\documentclass{ws-ijmpa-TOTEM}
\usepackage[super,compress]{cite}
\usepackage{graphicx}
\DeclareGraphicsExtensions{.ps,.eps,.pdf,.png,.gif,.jpg,.mht}
\usepackage{float,tabls}
\usepackage{color}
\usepackage{amsmath,amssymb,wasysym}
\usepackage{booktabs}
\usepackage{multirow}

\usepackage[breaklinks]{hyperref}
\hypersetup{colorlinks,urlcolor=black,citecolor=black,linkcolor=black,filecolor=black}
\newcommand{\rate}[2]{\ensuremath{N^{\text{#1}}_{\text{#2}}}}
\newcommand{\difrate}[2]{\ensuremath{\frac{\text{d}\rate{#1}{#2}}{\text{d}t}}}

\begin{document}
\markboth{TOTEM Collaboration }
{TOTEM Detectors Performance }

\renewcommand{\thefootnote}{\alph{footnote}} %
\title{ PERFORMANCE OF THE TOTEM DETECTORS AT THE LHC}

\author{G.~ANTCHEV,$^{a}$
P.~ASPELL,$^{8}$ 
I.~ATANASSOV,$^{8,a}$
V.~AVATI,$^{8}$ 
J.~BAECHLER,$^{8}$ 
M.~G.~BAGLIESI,$^{7b}$
V.~BERARDI,$^{5b,5a}$ 
M.~BERRETTI,$^{7b}$ 
E.~BOSSINI,$^{7b}$ 
U.~BOTTIGLI,$^{7b}$ 
M.~BOZZO,$^{6b,6a}$ 
E.~BR\"{U}CKEN,$^{3a,3b}$ 
A.~BUZZO,$^{6a}$ 
F.~S.~CAFAGNA,$^{5a}$
M.~G.~CATANESI,$^{5a}$
R.~CECCHI,$^{7b}$
C.~COVAULT,$^{9}$ 
M.~CSAN\'{A}D,$^{4,e}$
T.~CS\"{O}RG\H{O},$^{4}$ 
M.~DEILE,$^{8}$ 
M.~DOUBEK,$^{1b}$ 
K.~EGGERT,$^{9}$ 
V.~EREMIN,$^{b}$  
F.~FERRO,$^{6a}$ 
A. FIERGOLSKI,$^{5a,c}$ 
F.~GARCIA,$^{3a}$ 
S.~GIANI,$^{8}$ 
V.~GRECO,$^{7b}$ 
L.~GRZANKA,$^{8,d}$ 
J.~HEINO,$^{3a}$ 
T.~HILDEN,$^{3a,3b}$ 
A.~KAREV,$^{8}$
J.~KA\v{S}PAR,$^{1a,8}$ 
J.~KOPAL,$^{1a,8}$ 
V.~KUNDR\'{A}T,$^{1a}$ 
S.~LAMI,$^{7a}$ 
G.~LATINO,$^{7b}$ 
R.~LAUHAKANGAS,$^{3a}$ 
T.~LESZKO,$^{c}$ 
E.~LIPPMAA,$^{2}$ 
J.~LIPPMAA,$^{2}$ 
M.~LOKAJ\'{I}\v{C}EK,$^{1a}$
L.~LOSURDO,$^{7b}$  
M.~LO~VETERE,$^{6b,6a}$ 
F.~LUCAS~RODR\'{I}GUEZ,$^{8}$ 
M.~MACR\'{I},$^{6a}$ 
T.~M\"AKI,$^{3a}$ 
A.~MERCADANTE,$^{5a}$ 
N.~MINAFRA,$^{5b,8}$ 
S.~MINUTOLI,$^{6a,8}$ 
F.~NEMES,$^{4,e}$ 
H.~NIEWIADOMSKI,$^{8}$ 
E.~OLIVERI,$^{7b}$ 
F.~OLJEMARK,$^{3a,3b}$ 
R.~ORAVA,$^{3a,3b}$ 
M.~ORIUNNO,$^{f}$ 
K.~\"{O}STERBERG,$^{3a,3b}$ 
P.~PALAZZI,$^{7b}$ 
E.~PEDRESCHI,$^{7a}$
J.~PROCH\'{A}ZKA,$^{1a}$ 
M.~QUINTO,$^{5a,5b}$ 
E.~RADERMACHER,$^{8}$ 
E.~RADICIONI,$^{5a}$ 
F.~RAVOTTI,$^{8}$ 
E.~ROBUTTI,$^{6a}$ 
L.~ROPELEWSKI,$^{8}$ 
G.~RUGGIERO,$^{8}$ 
H.~SAARIKKO,$^{3a,3b}$ 
A.~SCRIBANO,$^{7b}$ 
J.~SMAJEK,$^{8}$ 
W.~SNOEYS,$^{8}$ 
F.~SPINELLA,$^{7a}$
J.~SZIKLAI,$^{4}$ 
C.~TAYLOR,$^{9}$
A.~THYS,$^{7b}$ 
N.~TURINI,$^{7b}$ 
V.~VACEK,$^{1b}$ 
M.~V\'ITEK,$^{1b}$ 
J.~WELTI,$^{3a,3b}$ 
J.~WHITMORE,$^{h}$ 
P.~WYSZKOWSKI$^{8,g}$
}

\address{$^{ 1a }$ Institute of Physics of the Academy of Sciences of the Czech Republic, Praha, Czech Republic.\\
$^{ 1b }$ Czech Technical University, Praha, Czech Republic.\\
$^{ 2 }$ National Institute of Chemical Physics and Biophysics NICPB, Tallinn, Estonia.\\
$^{ 3a }$ Helsinki Institute of Physics, Helsinki, Finland.\\
$^{ 3b }$ Department of Physics,  University of Helsinki, Helsinki, Finland.\\
$^{ 4 }$ MTA Wigner Research Center,  RMKI Budapest, Hungary.\\
$^{ 5a }$ INFN Sezione di Bari, Bari, Italy.\\
$^{ 5b }$ Dipartimento Interateneo di Fisica di  Bari, Italy.\\
$^{ 6a }$ Universit\`{a} degli Studi di Genova,  Genova, Italy.\\
$^{ 6b }$ Sezione INFN di Genova, Genova, Italy.\\
$^{ 7a }$ INFN Sezione di Pisa, Pisa, Italy.\\
$^{ 7b }$ Universit\`{a} degli Studi di Siena and Gruppo Collegato INFN di Siena,  Siena, Italy.\\
$^{ 8 }$ CERN,   Geneva, Switzerland.\\
$^{ 9 }$ Case Western Reserve University,  Dept. of Physics, Cleveland, OH, USA.\\
}

\maketitle

\begin{abstract}
The TOTEM Experiment is designed to measure the total proton-proton cross-section with the luminosity-independent method and to study elastic and diffractive pp scattering at the LHC. To achieve optimum forward coverage for charged particles emitted by the pp collisions in the interaction point IP5, two tracking telescopes, T1 and T2, are installed on each side of the IP in the pseudorapidity region  $3.1\leq \vert \eta \vert \leq 6.5$, and special movable beam-pipe insertions -- called Roman Pots (“RP”) -- are placed at distances of $\pm 147\:$m and $\pm 220\:$m from IP5. This article describes in detail the working of the TOTEM detector to produce physics results in the first three years of operation and data taking at the LHC.
\end{abstract}

\vfill\eject

\let\thefootnote\relax\footnote{$^a$ INRNE-BAS, Institute for Nuclear Research and Nuclear Energy, Bulgarian Academy of Sciences, Sofia, Bulgaria.}%
\footnote{$^b$ Ioffe Physical - Technical Institute of Russian Academy of Sciences, St.Petersburg, Russia.} %
\footnote{$^c$ Warsaw University of Technology, Warsaw, Poland.}%
\footnote{$^d$ Institute of Nuclear Physics, Polish Academy of Science, Cracow, Poland.}%
\footnote{$^e$ Department of Atomic Physics, E\"otv\"os University,  Budapest, Hungary.}%
\footnote{$^f$ SLAC National Accelerator Laboratory, Stanford CA, USA.}%
\footnote{$^g$ AGH University of Science and Technology, Krakow, Poland.}%
\footnote{$^h$ Penn State University, Dept.~of Physics, University Park, PA USA.}%
\addtocounter{footnote}{-8}
\newcommand{\thefootnote}{\alph{footnote}}


\ccode{PACS numbers: 13.85.-t Hadron-induced high- and super-high-energy interactions (energy $>$10~GeV), 13.85.Lg (Total Cross section), 13.85.Dz (Elastic scattering), 29.40.Gx (Tracking and position-sensitive detectors), 07.05.Hd (Data acquisition: hardware and software).}

\section{Overview}

The TOTEM experiment\cite{TOTEM-TDR} was specifically designed to measure the total proton-proton cross-section with the luminosity-independent method, theoretically based on the Optical Theorem, which requires the separate measurement of the elastic and inelastic cross sections.

Furthermore, TOTEM's physics programme aims at a deeper understanding of the proton structure by studying elastic scattering with large momentum transfers, and a comprehensive menu of diffractive processes partly in cooperation with the CMS experiment, located at the same interaction point IP5~\cite{TOTEM-CMS-diff-WG}. 

The experiment has been already described in detail in Ref.~\citen{Anelli:2008zza}.
This paper, after a brief overview of the experiment reviews specific issues related to the performance of the experimental apparatus and details the methods employed to calibrate the detectors and control the systematic of the data to produce physics results.\cite{totem1,totem6}

Two sets of  “Roman Pots” (RP) stations, placed at $\pm 147\:$m and $\pm 220\:$m from the interaction point, and their detectors, special silicon sensors  designed by TOTEM~\cite{Ruggiero:2005yd}, allow a detailed study of the elastic scattering cross-section down to a  four-momentum transfer squared of $|t|\approx 10^{-3}$ GeV$^2$.
To measure protons at the lowest possible emission angles, special beam optics have been implemented to optimise proton detection in terms of acceptance and resolution.

\begin{figure}[ht]
\centering
\includegraphics[width=0.99\linewidth]{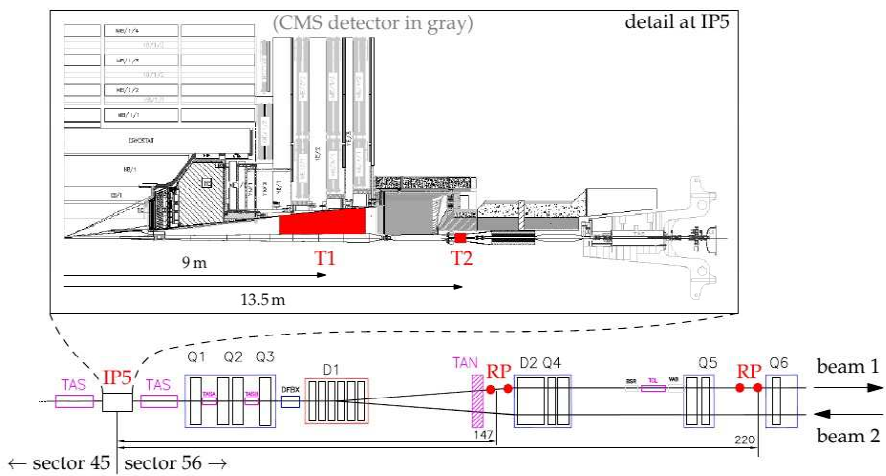}
\caption{A schematic view of the TOTEM detectors in the LHC.}
\label{fig:TOTEM-overall}
\end{figure}

To measure the inelastic cross-section by identifying the inelastic beam-beam events two telescopes (T1 and T2) detect charged particles produced in the pseudo-rapidity range of $3.1\leq \vert \: \eta \: \vert \leq 6.5$ ($\eta  = − \ln\, {\tan \theta }$). 
They provide a fully inclusive trigger for diffractive events and enable the reconstruction of the vertex of the interaction, in order to disentangle beam-beam events from the background. 
Each telescope is made of two arms, symmetrically placed at about 9 and 13.5 m from the IP respectively   (see Fig.~\ref{fig:TOTEM-overall}).

 The three TOTEM subsystems, being physically separate and different in design, have each their own particular electronic system, but nevertheless follow a common architecture. The read-out of all TOTEM detectors  is based on the custom-developed digital VFAT chip \cite{Kaplon:2005ce} that provides both tracking data and fast trigger signals.  
The data acquisition system is designed to be compatible with the CMS DAQ to make common data taking possible at a later stage.  
The TOTEM experiment can take data both in standalone mode or synchronised with CMS.

In view of a future common physics programme, and considering that the TOTEM detectors share their location with the CMS experiment, the TOTEM Software chain~\cite{totemsw} is written following the CMS Off-line Software framework (CMSSW~\cite{cmssw}) with its highly modular structure and the TOTEM related packages and data flow patterns can be easily incorporated into it.

To obtain physics quantities from  reconstruction of  the data collected in the experiment and detector simulation the off-line software uses as input both real and simulated data, along with meta-data, such as detector description, detector status, calibrations and alignment. 
The process, common to all three detectors, is organized in 3 steps: 
\begin{itemize}
\item {\it Cluster and Hit Reconstruction}: neighboring readout channels are converted into a single point (cluster); then converted to a Hit with coordinate (x,y,z) of the local detector reference system.

\item {\it Pattern Recognition and Local Reconstruction of the track}: grouping all the reconstructed points that belong to the same trajectory ("road search") and providing a track candidate for the fitting procedure after the detector alignment has been optimized with data.

\item {\it Global Track Reconstruction}: the reconstructed tracks are used to derive physics quantities, as the pseudorapidity (in the Telescopes) or the proton kinematics (in Roman Pots).
\end{itemize}

Detector specific procedures will be, where appropriate, detailed later when describing the performances of each detector. 

\section{ The Roman Pots: measurement of scattered protons}

To detect leading protons scattered at angles as small as 1$\mu$rad, silicon detectors are placed in the  “Roman Pots” (movable beam-pipe insertions) installed  symmetrically on either side of the LHC intersection point IP5. 
A RP {\it unit} consists of 3 RPs, two approaching the outgoing beam vertically and the third horizontally from the inside of the LHC ring: the detectors in the horizontal pot complete the acceptance for diffractively scattered protons. 
Two RP units {\it near} and {\it far}  separated by a distance of about $5\:$m  form a {\it station}. 

\subsection{The optics for the TOTEM specific runs}%

The angular range for detecting a leading proton in the detectors of the RP depends from the optics which is used to collide the beams in IP5.
Hence different optics allow TOTEM to measure elastically scattered protons over different angular ranges.
To access smaller $| t|$-values (at $\sqrt{s}= $7\,TeV, $|t| =$0.01\,GeV$^2$ corresponds to a scattering angle of $\approx 29 \mu$rad) the colliding beams must have a beam divergence of a few micro-radians (beam divergence $= \sqrt{ \epsilon /{\beta^∗}}$). 
This can be obtained by either increasing the betatron amplitude value at the interaction point ($\beta^*$) or by reducing the beam emittance $\epsilon$ .

An ultimate optics with $\beta^*$ of 1540 m was proposed and designed to eventually reach a region in four momentum transfer $t$ where the Coulomb and Nuclear scattering processes have similar amplitudes to allow for a precise determination of the parameter $\rho$, the ratio of the imaginary to the real part of the scattering amplitudes.

Measurements have been performed so far at the standard machine optics of $\beta^*$~=~0.6, 3.5, 11 m 6.5 and  7 m and with an optics with an intermediate  $\beta^*$ value of $90\,$m.  An optics with a value of $\beta^*$  of 1000\:m was tested and  used for a short period of data taking in 2012.
For the pA runs at the beginning of 2013 the machine optics was $\beta^*$~=~0.8 meters.

The $\beta^* = 90\, $m optics uses the standard injection optics and the beam conditions are typical for the early LHC running: zero degree crossing-angle and consequently at most 156 bunches with a low number of protons per bunch. 
The parallel-to-point focusing, which eliminates the dependence on the transverse position of
the proton at the collision point, is achieved in the vertical projection. 
At the same time, a large effective length ensures a sizable displacement from the beam centre.
Remarkably, the $\beta ^* = 90\,$m optics sensitivity to machine imperfections is sufficiently small from the viewpoint of data analysis \cite{Hubert-true}.

For a precise determination of the elastic scattering also the population of the bunches has to be controlled and kept below a value of $ N\approx 5\cdot 10^{10}$ p/bunch to minimize the correction due to multiple interactions in the same bunch crossing (overlapping events).

The RP detectors are housed, as said, in movable parts of the vacuum pipe: at injection when the entire machine aperture is needed by the machine the RP are moved in the garage position, i.e. out of the machine aperture. 
Once the beams are colliding and in stable conditions then the RP can be moved as close as possible to the beam.
Since the RP detectors are moved at every LHC fill also the position of the detectors with respect to the beam and to each other has to be checked for every run.
A specific procedure, which will be detailed later, has been developed to determine for each new optics the limiting position that the RP can be moved to.

\section{Tracking detectors for the RP}

\begin{figure}[htb]
\centering
\includegraphics[width=0.98\linewidth]{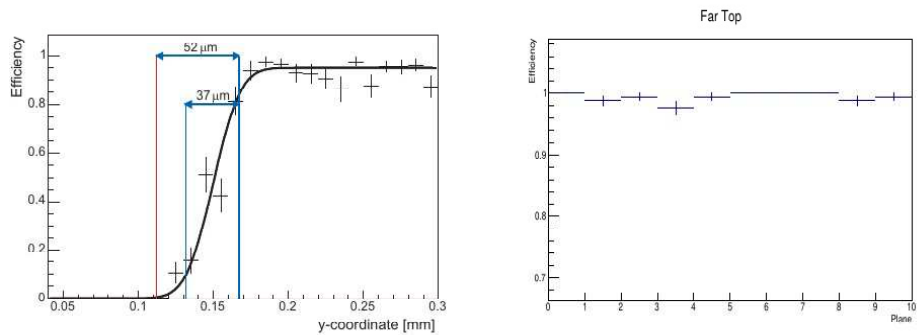} 
	\caption{Left plot: particle detection efficiency for a TOTEM silicon detector as a function of  the distance ($y$) from its mechanical edge. Right plot: hit efficiency of the 10 detectors of one pot for one of the $\beta^* =$ 90 m run (statistics: $~$2500 tracks). }
	\label{fig:edgeless-edge}
\end{figure}

Each RP is equipped with a stack of 10 silicon strip detectors designed with the specific objective of reducing the insensitive area at the silicon edge facing the beam.
The 512 strips with 66 $\mu$m pitch of each detector are oriented at an angle of $+ 45^o$ (five planes  of the stack) and $-45^o$  (five planes of the stack) with respect to the detector edge facing the beam. 
The detectors are able to trigger the data acquisition whenever a single proton goes trough one of the arms\footnote{Detectors in corresponding pots in a station (top, bottom or horizontal) form a telescope here named {\it arm}.}.
With carefully studied guard electrodes structure and employing specific dicing techniques the TOTEM silicon strip detectors allow a fully efficient particle detection already at a few tens of micrometers from the mechanical edge, as it can be seen in Fig.~\ref{fig:edgeless-edge} obtained during detector tests.
These studies are described in details in Ref.~\citen{Ruggiero:2009zz,Kaspar:2010zz,Edgeless-JINST-P06009,Hub-PhD}.

The very high efficiency of the silicon detectors in the experiment is shown in Fig.~\ref{fig:edgeless-edge}.

\begin{figure}[bt]
\centering
\includegraphics[width=0.49\linewidth]{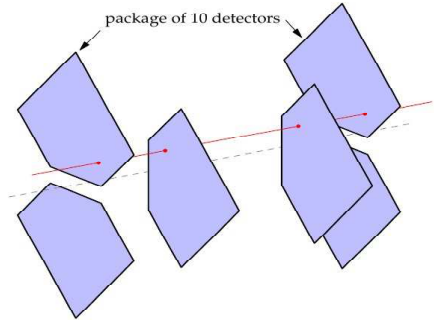}
\caption{Tracks transversing the overlap between vertical  and horizontal detectors establish the  alignment between the three mechanically independent detectors of a RP unit.}
\label{fig:stationscheme}
\end{figure}

The partial overlap between horizontal and vertical detectors (see Fig.~\ref{fig:stationscheme})  and the fact that the three RPs together with a Beam Position Monitor  (BPM) are rigidly fixed within a unit ensures the precision and the reproducibility of the alignment of all RP detector planes with respect to each other and to the position of the beam centre with a procedure that is detailed later (Section~\ref{sec:algnement}).

The 10 detector planes within a stack have been aligned and mounted with a precision of better than 20 $\mu$m and the three RPs and the BPM have been surveyed. 
The RP movement control system\cite{IPAC:2011} has been derived from the one of the LHC collimators.
The movements of the RPs via step motors (5  $\mu$m step) are independently verified with displacement inductive sensors (LVDT) with 10  $\mu$m precision. 

Proton kinematics at IP5 is reconstructed from positions and angles measured by the RP detectors, on the basis of the transport matrix between IP5 and the RP locations. 
The precision of optics determination is therefore of key importance for the experiment. 

\subsection{Track reconstruction and efficiency}%

The RP local reconstruction program calculates for each event the proton trajectories from the strips hit in the silicon detectors. 
Hits are transformed into strip clusters, and with geometry information are then converted to coordinates of points in space.
The pattern recognition process associates the sensor hits with particle tracks, and recognizes hits due to noise that are not associated with any track. 
The road search algorithm finds candidate tracks, which are approximately parallel to the beam, in a region which is $\approx 200\, \mu$m $\times 7 \,$milliradians wide.
In order to optimize the performance of the software module, only events with a signal in at least 3 planes per projection and with a number of hits between 1 and 5 per plane are considered.
 
\begin{figure}[ht]
\begin{center}
\includegraphics[width=0.49\linewidth]{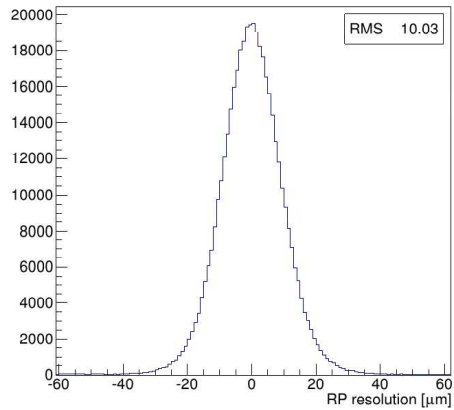}
\caption{One arm spatial resolution;  average resolution of the RP detectors of a near and far station; data from $\beta^*=90\,$m.}
\label{RP:spatial-res}
\end{center}
\end{figure}

The RP track candidates in the near and far units are then fitted with straight lines, whose parameters constitute the input to the proton reconstruction modules.
Fig.~\ref{RP:spatial-res} shows the spatial single arm resolution of the RP.

Track reconstruction may fail due to several reasons: intrinsic detection inefficiency of each silicon sensor, proton interaction with the material of a RP\footnote{protons to enter the detectors must traverse the thin stainless steel window 300$\mu$m thick that separates the LHC primary vacuum from the secondary RP vacuum.} and, the simultaneous presence of a beam halo particle and "pile-up" due to multiple interactions in the same bunch crossing. 
Besides detectors in the Roman Pot have the strips oriented in two directions and, while this allows for a good rejection of inclined background tracks, the possibility of identifying  more than one proton track almost parallel to the beam direction is very limited. 
Moreover protons cannot be reconstructed if they produced showers in the thin windows (0.3 mm) of either pot that separate the detectors from the machine vacuum.
These uncorrelated inefficiencies of single RPs are evaluated directly from the data. 
For example in case of an elastiuc scattering selection the result is a measured inefficiency of ($1.5 \pm  0.2$)\% for the near and ($3 \pm 0.2$)\% for the far RPs.
The different values can be explained by proton interactions in the near pot that affect the far RP too. 
This near-far correlated inefficiency is determined from data by counting events with corresponding shower signatures, yielding ($1.5 \pm 0.7$)\% (this result is confirmed by MC simulations).

\begin{figure}[hbt] 
\centering
\includegraphics[width=0.98\linewidth]{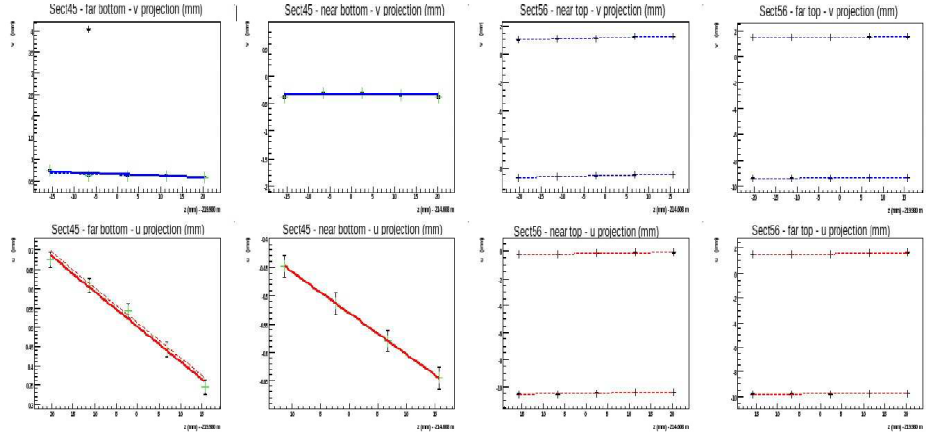} 
\caption{Overlapping events: the simultaneous presence of an elastic proton and a beam-halo proton. In one of the projections the tracks are almost overlapping making difficult to associate the hit to one of the two tracks.}
\label{fig:halo-and-proton}
\end{figure}

The most important contribution to the category of overlapping events is the simultaneous presence of an elastic proton and a beam-halo proton, see Fig. ~\ref{fig:halo-and-proton}.

The "pile-up" inefficiency is calculated from the probability of finding an additional track in any station of a "diagonal"\footnote{the set of RP stations on both sides of the IP that will detect an elastic event (i.e. for example  bottom detectors on the left (or bottom arm)  of the IP and top detectors (or top arm) on the right.)}. 
This probability is determined using a zero-bias data set.
Comparison of results from data sets obtained in different conditions shows that this inefficiency increases as RPs position gets closer to the beam. 
For the diagonal bottom-left top-right these probabilities are ($3.9 \pm 0.3$)\%, ($6.2 \pm 0.3$)\% and ($7.9 \pm 0.3$)\% for the datasets taken when the minimum distance to the beam was 6.5$\sigma$, 5.5$\sigma$, 4.8$\sigma$ respectively\footnote{$\sigma$ is the normalised beam dimension.}.

An average inefficiency of 3-7\% per pot and tracks induced correlations lead to an elastic event reconstruction inefficiency as determined  for each run directly from the data of (12.6 to 20.1)\%  at 7 TeV  and of (11.7 to 19.6)\% at 8 TeV.
A ranges for the values is due to different datasets (at different RP approaches) and considering both diagonals. 

\subsection{Proton reconstruction} 
\label{sec:RP:prot-reco}

Scattered protons are detected in the Roman Pots after having moved through a segment of the LHC lattice that contains 29 magnets per beam.
The trajectory of protons with transverse positions\footnote{The $*$ superscript indicates the LHC Interaction Point 5} $(x^*,y^*)$ and angles $(\Theta_x^*, \Theta_y^*)$ at IP5 are described with a linear formula  
\begin{align}
	\vec{d}=T\cdot\vec{d}^{*},  
	\label{proton_trajectories}
\end{align} 
where $\vec{d}=\left\{x,\Theta_x,y,\Theta_y,\Delta p/p\right\}^{T}$ with nominal beam momentum $p$  and momentum loss $\Delta p$. 

The transport matrix  $T\left(s;\mathcal{M}\right)$ is defined by the optical functions
\begin{align}
	T=\left(
		\begin{array}{ccccc} 
			v_x         & L_x     & m_{13}   & m_{14}  & D_x  \\
			v_x'        & L_x'    & m_{23}   & m_{24}  & D_x' \\
			m_{31}      & m_{32}  & v_y      & L_y     & D_y  \\
			m_{41}      & m_{42}  & v_y'     & L_y'    & D_y' \\ 
			0           & 0       & 0        & 0       & 1
    	\end{array}  
	\right).
	\label{transport_matrix}
\end{align}
The \textit{magnification} $v_{x,y}=\sqrt{\beta_{x,y}/\beta^*}\cos\Delta\phi_{x,y}$ and the  {\it effective length} $L_{x,y}=\sqrt{\beta_{x,y}\beta^*}\sin\Delta\phi_{x,y}$ are functions of the {\it betatron amplitude} $\beta_{x,y}$ and the relative {\it phase advance} $\Delta\phi_{x,y}=\int^{\text{\tiny RP}}_{\text{\tiny IP}}\beta(s)_{x,y}^{-1}ds$ and are particularly important for the proton kinematics reconstruction. 
The coupling coefficients $m_{i,j}$ are close to 0 and the vertex contributions cancel due to the symmetry of the scattering angles.  
Therefore, the kinematics of elastically scattered protons at IP5 can be reconstructed from Equation~(\ref{proton_trajectories}) as:
\begin{align}
	\Theta_{y}^{*} \approx \frac{y_{\text{\tiny RP}}}{L_{y,\text{\tiny RP}}}\;\;\;\;\;\;
	\Theta_{x}^{*} \approx \frac{1}{\frac{dL_{x,\text{\tiny RP}}}{ds}}\left(\Theta_{x,\text{\tiny RP}}-\frac{dv_{x,\text{\tiny RP}}}{ds}x^{*}\right),
\end{align}
where the subscript ``RP" indicates the value at the measurement location. As the values of the reconstructed angles are inversely proportional to the optical functions, the knowledge of the accuracy of the optics determines the systematic errors of the final physics results.\par

\subsection{Optics determination with proton tracks}
\label{sec:RP:optics-det}

TOTEM developed a novel method of machine optics determination making use of angle-position distributions of elastically scattered protons observed in the RP detectors together with the data retrieved from several machine databases. 
Studies show that the transport matrix could be estimated with a precision better than 1\%.

This method has been successfully applied to the data taken so far. 

The proton transport matrix $T\left(s;\mathcal{M}\right)$ over a distance $s$  is defined by the machine settings $\mathcal{M}$. 
It is calculated with the MAD-X \cite{Grote:2003ct} code for each group of runs with identical optics using values logged in several data sources: the magnet currents are retrieved from TIMBER \cite{TIMBER:2006} and 
the WISE database \cite{WISE} and include measured imperfections (field harmonics, magnet displacements and rotations).\par
However, the lattice is subject to additional imperfections ($\Delta \mathcal{M}$) not known with enough precision, which alter the transport matrix by $\Delta T$:
\begin{align}
	T\left.(s;\: \mathcal{M}\right) \rightarrow T\left.(s;\: \mathcal{M}+\Delta \mathcal{M}\right) = T\left.(s;\: \mathcal{M}\right)+\Delta T .
\label{Transport-corrections}
\end{align}
The 5--10\% precision of the $\Delta \beta/\beta$ ($\beta$ beat) measurement does not allow to estimate $\Delta T$ with the accuracy required by the TOTEM physics program.
However, the magnitude of $\left|\Delta T\right|$ can be evaluated from the tolerances of the LHC imperfections of which the most important are:
\begin{itemize}
	\item Magnetic field strength conversion error $I \rightarrow B\:, \sigma(B)/B\approx10^{-3}$
	\item Beam momentum offset $\sigma(p)/p \approx 10^{-3}\:$.
\end{itemize}
Their impact on optical functions is presented in Table~\ref{tbl:Lysensitivity}. 
It is clearly visible that the imperfections of the inner triplet (the MQXA and MQXB magnets) have a sizable impact on the transport matrix while the optics is less sensitive to the quadrupoles MQY and MQML.
 
Other imperfections of lower significance are:
\begin{itemize}
	\item Magnet rotations $\sigma(\phi)\approx1$ mrad
	\item Beam harmonics $\sigma(B)/B \approx 10^{-4}$
	\item Power converter errors $\sigma(I)/I \approx 10^{-4}$
	\item Magnet positions $\Delta x,\Delta y \approx 100\:\mu$m.
\end{itemize}
Generally, as can be seen in Table~\ref{tbl:Lysensitivity}, for large $\beta^*$ optics the magnitude of $\Delta T$ is sufficiently small from the viewpoint of data analysis and therefore $\Delta T$ does not need to be  estimated with more precision. However, the small $\beta^*$ optics sensitivity to the machine imperfections is significant and cannot be neglected. Fortunately, in this case $\Delta T$ can be determined precisely enough from the proton tracks in the Roman Pots.

\begin{table}[hbt] 
 \tbl{Sensitivity of the vertical effective length $L_y$ to magnet strengths and beam momentum perturbed by 1~$\permil$ for low- and large-$\beta^*$ optics. }
      { \begin{tabular}{ p{5cm} |  p{3cm} | p{3cm} |} 
\cline{2-3}           									
		& \multicolumn{2}{|c|}{\bf $\mathbf{\boldsymbol\delta L_y/L_y}$\:[\%]}      \\ 
\hline
   \multicolumn{1}{|c|}{\bf Perturbed element}   	&  \multicolumn{1}{|c|}  {\hspace{0.8cm}$\mathbf{\boldsymbol\beta^{*}=3.5\:m}$\hspace{0.8cm} } 	&  \multicolumn{1}{|c|}{\hspace{0.8cm}$\mathbf{\boldsymbol\beta^{*}=90\:m}$ \hspace{0.8cm} }	\\ 
\hline
            \multicolumn{1}{|c|}{MQXA.1R5}        		& \multicolumn{1}{|c|}{ $\phantom-0.98$ }           				&  \multicolumn{1}{|c|}{$\phantom-0.14$ }   						\\
            \multicolumn{1}{|c|}{MQXB.A2R5}       		& \multicolumn{1}{|c|}{ $-2.24$  }          					&  \multicolumn{1}{|c|}{$-0.23$}    						\\
            \multicolumn{1}{|c|}{MQXB.B2R5}       		& \multicolumn{1}{|c|}{ $-2.42$}           					& \multicolumn{1}{|c|}{ $-0.25$ }\\
	            \multicolumn{1}{|c|}{MQXA.3R5}        		&  \multicolumn{1}{|c|}{$\phantom-1.45$ } 	     			&  \multicolumn{1}{|c|}{ $\phantom-0.20$}    						\\
            \multicolumn{1}{|c|}{MQY.4R5.B1}      		&  \multicolumn{1}{|c|}{$-0.10$ }          	  				& \multicolumn{1}{|c|}{ $-0.01$}    						\\
            \multicolumn{1}{|c|}{MQML.5R5.B1}     		& \multicolumn{1}{|c|}{ $\phantom-0.05$ }         				&  \multicolumn{1}{|c|}{$\phantom-0.04$}	\\
            \multicolumn{1}{|c|}{$\Delta$p/p}     		& \multicolumn{1}{|c|}{ $-2.19$  }           					&  \multicolumn{1}{|c|}{$\phantom-0.01$ }	\\ \hline
       \end{tabular} \label{tbl:Lysensitivity} }
\end{table}

\subsubsection{Constraining the parameters of the optics with the proton tracks}

The RP detector system, due to its high resolution ($\sigma(x,y) \approx 11\,\mu$m, $\sigma(\Theta_{x,y}) \approx 2.9\,\mu$rad), can measure very precisely the proton angles, positions and the angle-position relations on an event per event basis. 
These quantities can be employed to define a set of estimators characterising the correlations between the elements of the transport matrix $T$ or between the transport matrices of the two LHC beams. 
Such a set of estimators $\hat{R}_1, ..., \hat{R}_{10}$ (defined in the subsequent sections) is exploited to reconstruct, for both LHC beams, the imperfect transport matrix $T(\mathcal{M})+\Delta T$ defined in Equation~\ref{Transport-corrections}.

The elements of the transport matrix are functions of the betatron amplitudes $\beta_{x,y}$ and the phase advances $\phi_{x,y}$ and therefore they are mutually related. 
Since the momentum of the two LHC beams is identical, the elastically scattered protons will be deviated symmetrically from their nominal trajectories of Beam 1 and Beam 2: 
\begin{align}
    \Theta^{*}_{x,b_1} = \Theta^{*}_{x,b_2}\:,\; 
    \Theta^{*}_{y,b_1} = \Theta^{*}_{y,b_2}\:, 
    \label{collinearity_cut}
\end{align}
which allows to compute ratios between the effective lengths of the two beams. From Equation (\ref{proton_trajectories}) we get: 
\begin{align}
    R_{1}&\equiv\frac{\Theta_{x,b_1,\text{RP}}}{\Theta_{x,b_2,\text{RP}}} \approx \frac{\frac{dL_{x,b_1,\text{RP}}}{ds}\Theta^{*}_{x,b_1}}{\frac{dL_{x,b_2,\text{RP}}}{ds}\Theta^{*}_{x,b_2}} =         \frac{\frac{dL_{x,b_1,\text{RP}}}{ds}}{\frac{dL_{x,b_2,\text{RP}}}{ds}}\:, 
    \label{ratiodLxds}
\end{align}
\begin{align}    R_{2}&\equiv\frac{y_{b_1,\text{RP}}}{y_{b_2,\text{RP}}} \approx \frac{L_{y,b_1,\text{RP}}}{L_{y,b_2,\text{RP}}},
\label{ratioLy}
\end{align}
where $b_1$ and $b_2$ indicate beam 1 and beam 2. 
 Approximations present in Equations~(\ref{ratiodLxds}) and~(\ref{ratioLy}) represent the impact of statistical effects such as detector resolution, beam divergence and primary vertex position distribution. 
The estimators $\hat{R_1}$ and $\hat{R_2}$ are finally obtained from the $(\Theta_{x,b_1}, \Theta_{x,b_2})$ and $(y_{b_1,220},y_{b_2,220})$ distributions and are defined with the help of their principal eigenvectors. 
A precision of 0.5\% is attained. 

Furthermore, from the distributions of proton angles and positions for elastically scattered protons detected in Roman Pots, one can also measure ratios of other elements of the transport matrix $T$. 
First of all, $dL_y/ds$ and $L_y$ are related by 
\begin{align}
	R_3\equiv\frac{\Theta_{y,b_1,\text{RP}}}{y_{b_1,\text{RP}}} \approx \frac{\frac{dL_{y,b_1,\text{RP}}}{ds}}{L_{y,b_1,\text{RP}}}\:,\;
\end{align} 
An analogous value $R_4$ is defined for beam 2.
The corresponding estimators $\hat{R_3}$ and $\hat{R_4}$ can be calculated with a precision of 0.5\% from distributions .

Similarly, we exploit the horizontal distributions to quantify the relation between $dL_x/ds$ and $L_x$.
However when for the optics used by TOTEM  $L_x$ is close to $0$, instead of defining the ratio we rather estimate the position $s$ (with the precision of about $1\:$m) along the beam where $L_x$ equals to $0$ by solving 
\begin{align}
	\frac{L_x(s)}{dL_x(s_1)/ds}= \frac{L_x(s_1)}{dL_x(s_1)/ds} + \left(s-s_1\right)=0\:,
\end{align}
where $s_1$ is the position of the Roman Pot station. The ratio $\frac{dL_{x}(s_1)}{ds}/L_{x}(s_1)$ is extracted by the proton distributions.  

Finally, tracks determine as well the coupling components of $T$.

\subsubsection{Optics matching}
On the basis of the constraints described above, 
$\Delta T$ can be determined with a $\chi^2$ minimization procedure. 
The selected lattice imperfections  form a 26 dimensional optimization phase space, which includes the magnet strengths, rotations and beam momenta. 
Due to the high dimensionality of the phase space and the approximate linear structure of the problem there is no unique solution. The result of the optimization depends also on additional constraints imposed by the machine tolerances.  The  $\chi^2$ is composed of the part defined by the values measured with the Roman Pots (discussed in the previous section) and the ones reflecting the LHC tolerances:
\begin{align}
\chi^2 = \chi_\text{Measured}^2 + \chi_\text{Design}^2,
\end{align}
where the design part  
defines the nominal machine as an attractor in the phase space, and the measured part
contains the track based constraints together with their errors optimized with the MAD-X software. 

Table \ref{tbl:matching_result} presents the results of the optimization procedure for $\beta^*=3.5\:$m. The  value obtained for the effective length $L_y$ of beam 1 is close to nominal, while for beam 2 there is a significant variation. The same pattern applies to the values of $dL_x/ds$. 

\begin{table}[hbt]
\renewcommand{\arraystretch}{1.1}\addtolength{\tabcolsep}{-4pt}
\tbl{Comparison between  LHC beams optical functions and their nominal values for $\beta^*=3.5\:$m.}
{\begin{tabular}{ c |c|c|c|c|}\cline{2-5}
	 							&	$\mathbf{L_{y,b_1}}$[m]  		&	$\mathbf{dL_{x,b_1}/ds}$		& {\bf $\mathbf{L_{y,b_2}}$[m]} 	&   $\mathbf{dL_{x,b_2}/ds}$	\\\hline	
	 \multicolumn{1}{|c|}{\bf Nominal}		& 	$22.4$ 					&	$-0.321$ 					& $18.4$ 						&	$-0.329$ 	\\\hline
	 \multicolumn{1}{|c|}{ \bf Estimated}	& 	$22.6$					& 	$-0.312$ 					& $20.7$						& 	$-0.315$	\\\hline
	\end{tabular} \label{tbl:matching_result} }
\end{table}

The procedure has been extensively verified with Monte Carlo studies. The nominal machine settings were perturbed in order to simulate the LHC imperfections and then the simulated proton tracks were used to calculate the constraints $R_1$ to $R_{10}$. The study included the effect of: 
	\begin{itemize}
		\item magnet strengths 
		\item beam momenta
		\item displacements, rotations
		\item kickers, harmonics
		\item elastic scattering $\Theta$-distributions
	\end{itemize}\par
\begin{figure}[htb]
\centering
\includegraphics[width=0.98\linewidth]{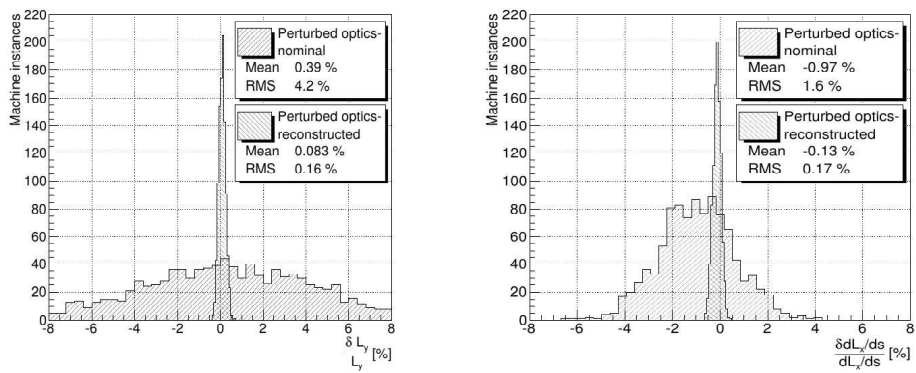} 
	\caption{Relative error distribution of $L_{y}$ for beam 1 before and after matching (left) and of $dL_{x}/ds$ for beam 1 before and after matching(right).}
	\label{fig:MCresultLy}
\end{figure}

The results obtained from the study of the $\beta^*=3.5\:$m optics are summarized in Fig. \ref{fig:MCresultLy}  and their statistical relevance is given in Table \ref{tbl:MCestimations}. 

The distributions of optical functions' errors indicate that the optical functions can be reconstructed with a precision of 0.2\%, which confirms the validity of the proposed approach.

\begin{table}[hbt]
\renewcommand{\arraystretch}{1.1}
\tbl{Monte-Carlo validation results of Roman Pot track based optics estimation. The machine imperfections induce large spread of optical functions. The matching procedure estimates the optics with errors lower than 2.1 $\permil$.} 
{\begin{tabular}{  c | c | c | c | c | c | c |} \cline{2-5}
       & \multicolumn{2}{|c|}{\bf Simulated} & \multicolumn{2}{|c|}{\bf Reconstructed} \\
       & \multicolumn{2}{|c|}{\bf optics distribution} & \multicolumn{2}{|c|}{\bf optics error} \\ \hline
       \multicolumn{1}{|c|}{ \bf Relative optics }                                      & { Mean}   & { RMS}        & { Mean}                   & { RMS}   \\
       \multicolumn{1}{|c|}{\bf distribution}                                           &    [\%]      &     [\%]       &     [\%]                      & [\%]  \\ \hline
       \multicolumn{1}{|c|}{ $\frac{\delta L_{y,b_{1}}}{L_{y,b_{1}}}\;$ }               &  $\phantom-0.39$         & $4.2$           & $\phantom-0.083$    & $0.16$ \\
       \multicolumn{1}{|c|}{ $\frac{\delta dL_{x,b_{1}}/ds}{dL_{x,b_{1}}/ds}\;$ }       &  $-0.97$          & $1.6$           & $-0.13$             & $0.17$  \\\hline
        \multicolumn{1}{|c|}{ $\frac{\delta L_{y,b_2} }{L_{y,b_2}} \;$ }            &  $-0.14$          & $4.9$           & $\phantom-0.21$    & $0.16$ \\
        \multicolumn{1}{|c|}{ $\frac{\delta dL_{x,b_2}/ds}{dL_{x,b_2}/ds}\;$ }      & $\phantom-0.10$         & $1.7$           & $-0.097$              & $0.17$  \\\hline

   \end{tabular} \label{tbl:MCestimations} }
\end{table}

It is foreseen to extend this approach to model also the transport of protons with large momentum losses. 

A more extensive and detailed description of this method can be found in Ref.~\citen{Hubert-true}.

\subsection{Alignment of RP, detectors and proton reconstruction}%

Precise knowledge of the position of the detector with respect to the circulating beam at the moment of the measurement (order of $\mu $m) is  needed for physics performance and, since one desires to approach the beam as closely as possible, a delicate procedure must be repeated for each run.
The alignment process can be seen as composed of three steps applied in sequence: first the detector housing (i.e. the RP) position with respect to the beam must be known as precisely as possible, second the individual detectors need to be aligned with respect to each other and third the global alignment of detector in different RPs is performed using elastic events.
These steps are described in what follows. 

\subsubsection{Alignment between the RP and the circulating beam}\label{sec:algnement}

\begin{figure}[hbt]
\centering
\includegraphics[width=0.98\linewidth]{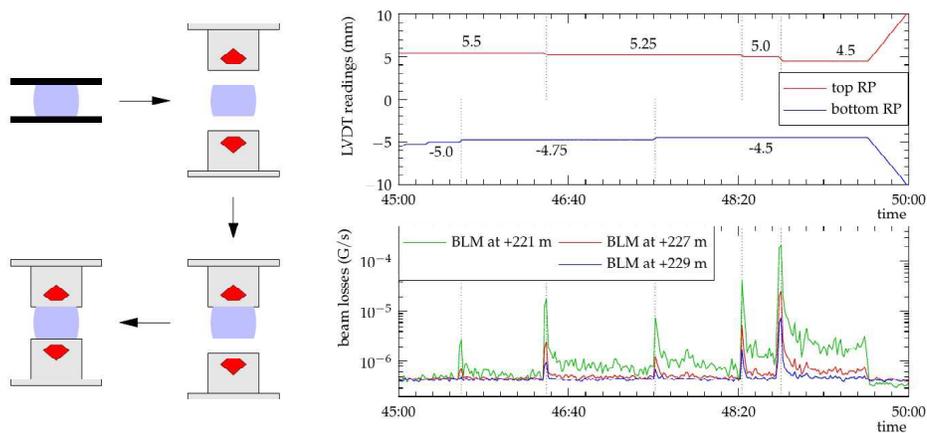}
\centering
\caption{Beam-based alignment sequence for a vertical RP pair in 2010. The left figure shows the sequence of steps and the right one the correlation between the RP movement towards the beam and the beam losses downstream of it (on the x-axis the time is indicated in minutes).
}
\label{fig:RP-align}
\end{figure}

A special fill is required to determine the position of the RPs  and their detectors (alignment) with respect to the circulating beam with a procedure similar to the one employed for the LHC collimators: 
\begin{itemize}
\item a collimator scrapes the beam at a distance from the beam center defined by the machine operation  (left plot in Fig. \ref{fig:RP-align}) creating a sharp beam edge. 
\item the RP are moved one at a time towards the recently created sharp beam edge until an increase (spike) in beam losses is recorded downstream of the RPs (right plot in Fig. \ref{fig:RP-align}). 
\end{itemize}
The RP window and the primary collimator that essentially defines locally the machine aperture are now at the same distance from the beam orbit: during data taking with more intense fills the RP are retracted with respect to this distance on closest approach. 
The distance from the beam is expressed with the normalised dimension of the beam ($\sigma_{beam}$) at the RP position.

The precision of this procedure is determined by the movement step size: in the early alignment exercises of 2010 (summarized in Fig. \ref{fig:RP-align}) the step size was large (250 $\mu$m) while more recently the step size varies between 10 and 50 $\mu$m.
Due to a poorly defined gap between the thin windows and the silicon sensors, the alignment between the sensors and the beam has an uncertainty of 200 $\mu $m.\cite{IPAC:2011}

\subsubsection{Track-based alignment}

The individual detector planes in a detector package and in a unit are aligned with respect to each other using reconstructed tracks.
The underlying idea is that sensor misalignments give rise to residuals, i.e. the distances of the measured hit positions from the fitted tracks, which need to be minimised in the process. 
This technique is sensitive to shifts and rotations of individual detector planes relative to each other, but not to global shifts or rotations which are determined using an elastic events subsample as described later. 

The transverse overlap between vertical  and horizontal detectors (Fig.~\ref{fig:stationscheme}) allows to measure precisely the  alignment between the three mechanically independent set of detectors in a RP unit.
The elastic scattering events that cross one of the vertical detectors above the beam on one side of the IP will cross the ones below the  beam on the other side and vice-versa.  
The two different subset are independent measurement of the same physics process and the distance between two opposed detectors enters in the determination of the value of t or the scattering angle. A very precise determination of this relative distance is important to reduce the systematic errors.

\begin{figure}[htb]
    \centering
\includegraphics[width=0.8\linewidth]{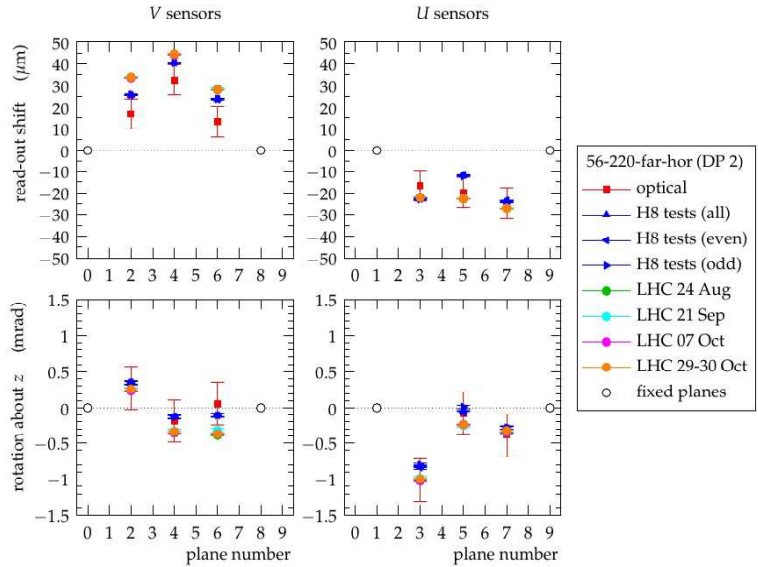}
    \caption{Comparison of different alignment result for one example detector 
package. ``H8'' refers to a test beam alignment before installation, ``LHC'' 
to an alignment with LHC physics data in 2010, ``optical'' to the 
metrology measurement during assembly.}
    \label{fig:trackbasedresult}
\end{figure}

The precision of the alignment procedure depends on the statistics and track distributions, but the typical uncertainties reached are few micrometres for the shifts and less than $0.1$ milliradians for the rotations.
Fig.~\ref{fig:trackbasedresult} shows a comparison of the track-based alignment results for one RP with the detector alignment obtained via optical metrology in the laboratory.

\subsubsection{Alignment using elastic events}

The global misalignment modes (e.g. common shifts or rotations of the entire unit w.r.t. the beam) are inaccessible to the track-based techniques but can be constrained by exploiting known symmetries of certain physics processes.
A prominent example is the hit distribution in the vertical detectors of elastic scattering events that allows, once the optics has been properly understood, to consider the two elastically scattered protons as an ideal ruler. %

\begin{figure}[hbt]
\centering
\includegraphics[width=0.98\linewidth]{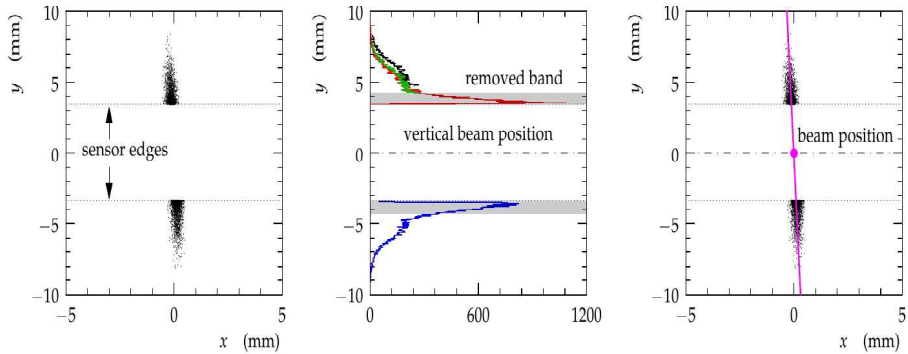}
\caption{Alignment with selected elastic events. Left plot:  a selection of elastic events; center plot: vertical alignment; right plot:  horizontal alignment and the unit rotation are established by fitting the track distribution.  The procedure is discussed in detail in the text.}
 \label{fig:elastic-based-align}
\end{figure}

The RP station alignment based on the symmetry of the elastic scattering hit distribution is illustrated in Fig.~\ref{fig:elastic-based-align}. 

The left-hand side plot shows a distribution of track intercepts for vertical detectors in a scoring plane of a RP station at 220 m  for a run with $\beta ^* = 3.5 \:$m at $\sqrt{s} = 7 \, $TeV.   
The tilt of the vertical band is mainly caused by optics imperfections.

The center plot shows the vertical alignment. 
Data that might be affected by acceptance effects (grey bands) are removed from the vertical hit distributions. 
The symmetry line of the vertical distribution is the position of the beam centre  (black dash-dotted line) and  is obtained by inverting the sign of the $y < 0$ distribution (blue), and shifting it until it coincides with the $y > 0$ part (red).

Finally the horizontal alignment and the unit rotation are established by fitting the track distribution, as shown in the right plot. 

The precision that can be obtained depends on the statistics and the optics parameters: for example for the  $\beta ^*= 90\,$m optics the precision is known to better than 5 $\mu $m (horizontal) and about 30 $\mu $m (vertical). 
The tilts of the detectors  are determined with an uncertainty about $0.1$ milliradian.

A detailed discussion of the alignment methods is given also in Ref.~\citen{jan-thesis}.

\subsection{ Acceptance and resolution}

The acceptance of the RP system for elastically or diffractively scattered protons depends on the optics configuration. 
The proton acceptance of a RP station is determined by the minimum distance of a RP device to the beam and by constraints imposed by the beam pipe or beam screen size.

\begin{figure}[ht]
\begin{center}
\includegraphics[width=0.98\linewidth]{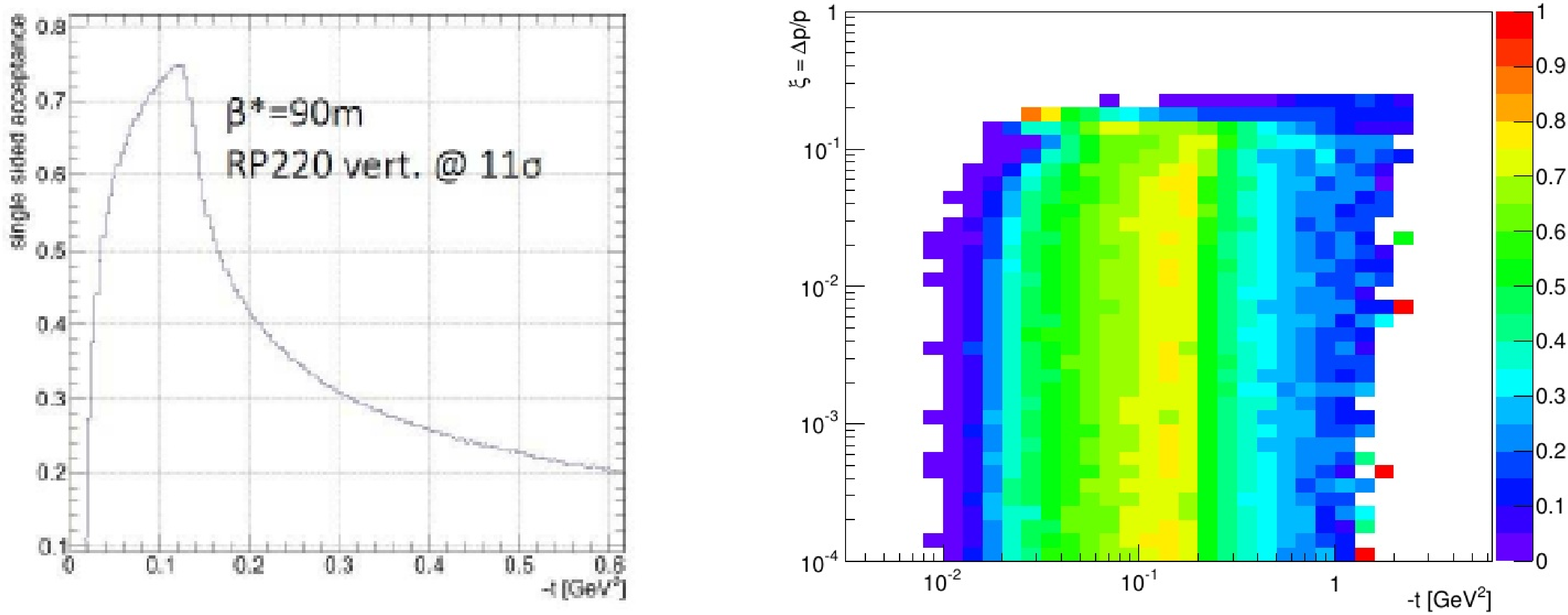}
\caption{The TOTEM acceptance as a function of the $|t|$ of the measured proton when the RP detectors are at a distance of 11 $\sigma$ from the circulating beam. }
\label{rp_acceptance}
\end{center}
\end{figure}

The minimum observable values of $|t|$ (for the vertical detectors) and $|\xi|$ (for the horizontal detectors) are given by the distances of the RPs from the beam centre. 
This distance is defined as a multiple $K$ of the beam width $\sigma_{x,y}$ and has lower limits determined by arguments of machine protection. 
For regular fills, $K\ge 11$; in special runs immediately after a beam-based alignment, approaches as close as $K = 3$ have been realised. 

\begin{align}
|t|_{min} = { {p^2 (K_{\sigma _y} + \delta)^2}\over { L^{2}_{y}}} \hspace{1cm}{\rm (vertical \; detectors)} \\
|\xi_{min}| ={{K_{σ_x} + \delta}\over {Dx}} \hspace{1cm}{\rm (horizontal \; detectors) .}
\label{equ:csi-min}
\end{align}

The active detector area starts at a typical distance $\delta = 0.5$mm from the surface of the RP window nearest to the circulating beam.
The acceptance of the RP220 station of elastic and diffractive protons for a   $\beta ^* = 90 \, $m optics is shown in Figure~\ref{rp_acceptance}.
For the $\beta^*$= 90 m optics, the vertical effective length $L_y$ at the RP220 station equals 260 m: if the vertical Roman Pots are placed 10 mm from the beam centre, the lowest detected $\Theta^*_y$ is 40 $\mu$rad and the t-acceptance starts at $2\times10^{-2} $ GeV$^2$.

The maximum of the acceptance is limited by the aperture of the accelerator\footnote{The size of the vacuum pipe at a specific location finally determines the largest angle accepted by a specific optics.}.
A value of the horizontal effective length $L_x$ close to 0 allows for detection of high values of $\Theta^*_x$ and an acceptance in $t$ for this optics up to 1 GeV$^2$.

\begin{figure}[ht]
\begin{center}
\includegraphics[width=0.49\linewidth]{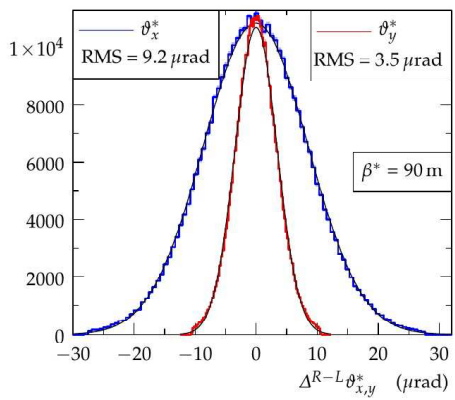}
\caption{ Single arm angular resolution $\theta^*_x$ and  $\theta^*_y$ extracted from elastic data .}
\label{RP:theta_res_elastic}
\end{center}
\end{figure}

Within the $t$-acceptance range, diffractively scattered protons are detected independently from their momentum loss, and thus the entire $\xi$-range can be observed.

The full set of kinematic variables is reconstructed with the use of the parameterised proton transport functions, discussed in  Sections \ref{sec:RP:prot-reco} and \ref{sec:RP:optics-det}.%
The details of the reconstruction algorithms and optics parameterization are discussed in\citen{Hub-PhD}. 
The scattering angle resolution depends mainly from the angular beam divergence and from the detector resolution.
\begin{figure}[ht]
\begin{center}
\includegraphics[width=0.98\linewidth]{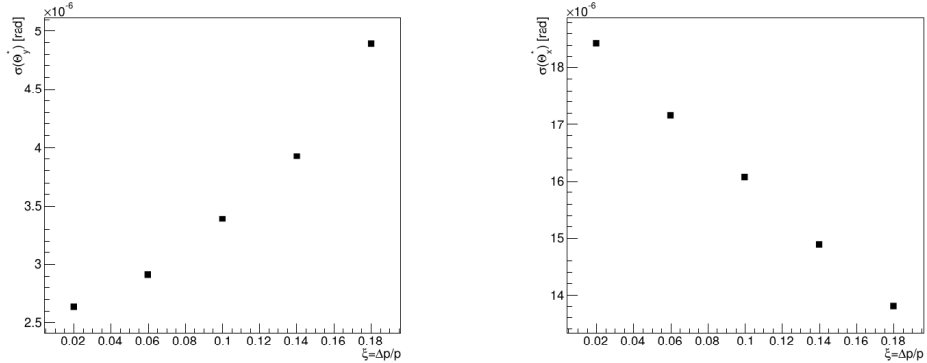}
\caption{Angular resolution for detection in the horizontal and vertical plane from a sample of diffractively scattered protons (the first bin contains data points  with $\xi\le0.02$) }
\label{RP:theta_res_diff}
\end{center}
\end{figure}
In case of elastic events the angular resolution is determined by comparing the scattering angles reconstructed from the left and right arm.  Fig.~\ref{RP:theta_res_elastic} shows one-arm resolutions obtained from elastic events.
The resolution improves by a factor $\sqrt{2}$ when the left and right arm measurements are averaged.
In the case of diffractively scattered protons the values of optical function parameters vary with the proton momentum loss $\xi$ and this results in $\xi$-dependent resolution for the main variables $\Theta_{x,y}^*$ and $\xi$.
\begin{figure}[t]
\begin{center}
\includegraphics[width=0.98\linewidth]{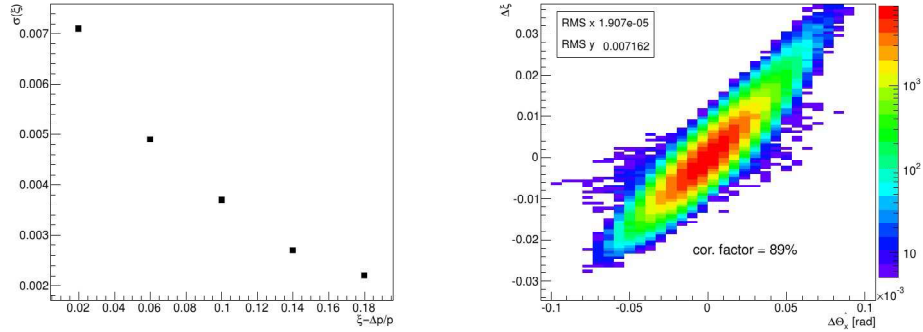}
\caption{Left plot: momentum reconstruction resolution for a diffractive protons sample. Right plot: correlation between the errors in $\Theta^*_x$ and $\xi$.}
\label{RP:xi-resolution}
\end{center}
\end{figure}
Moreover, as a dispersion term is present in the horizontal projection (cfr. eq.~\ref{equ:csi-min}), both $\xi$ and  $\Theta_{x}^{*}$ contribute to determine the horizontal trajectory resulting in a correlation of these two variables. 
The resolution of $\Theta_y^*$ and $\Theta_x^*$ reconstructed for diffractively scattered protons as a function of  $\xi$ is shown in Fig.~\ref{RP:theta_res_diff} and Fig.~\ref{RP:xi-resolution}.

Since for low $\xi$ the $L_x$ value is close to 0, the horizontal component of the scattering angle is reconstructed with the resolution of 20$\,\mu$rad, which is an order of magnitude worse than the beam divergence limit. 
However for the vertical plane $L_y$ is 260 m and the resolution for the reconstruction in y is 2.5 $\mu$rad, very close to the beam divergence limit.

 \subsection{The RP Trigger} %

The 10 detector planes (5 planes per each strip orientation u and v) are used to generate the trigger of one Roman Pot station. 
The VFAT R/O chip  generates a fast signal whenever a particle traverses one of a group of 32 adjacent channels.  
Each detector plane is divided in 16  groups of 32 strips (corresponding physically to a region of the detector 2\:mm wide)

Bits corresponding to the 32 detector regions (16 in u and 16 in v) are then analysed by the coincidence chip (CC) which defines a track road per strip orientation requiring a coincidence of at least 3 out of 5 planes.  
This loose requirement includes tracks crossing the boundary of neighbor track roads.
Coincidences between track segments in the u and v planes are used to further constrain the trigger rate. 

Signals are then sent to the counting room where a coincidence between track segments from two RP units in a station and the different stations is performed; multiplicity cuts reduce further the background originating from beam-gas interactions (see Section \ref{trigger-section}).

\section{The T2 telescope}
\label{T2-section}

TOTEM detects a very large fraction ($\approx 94\%$) of the inelastic cross section with the very forward angular coverage provided by T2.\cite{totem5}%

The T2 telescope, placed at about 14 meters from the interaction point (IP), covers the pseudorapidity region between 5.3$\,< |\eta| <\,$6.5. 
There are two T2 telescopes, one on each side of the IP, both consisting of 2 quarters with 10 detectors.
The detectors of the T2 telescope are semicircular triple-GEM (Gas Electron Multipliers)\cite{SAULI}, gas-filled detectors that have the advantage of separating the charge amplification structure from the charge collection and readout structure.
The T2 GEM detectors have an almost semi-circular shape\cite{GEM:construction} and are assembled  to surround the LHC vacuum pipe with seamless $\phi$ coverage.
Each detector provides a two-dimensional information of the track position in an azimuthal coverage of \(192\,^{\circ}\), which allows an overlap region along the vertical axis between two neighboring half-telescopes\footnote{since there are four of these half telescopes in the experiment these are often referred to as "quarters".} quarters.

The read-out board of the GEM detectors is double layered and contains two columns of 256 concentric strips (400$ \, \mu$m pitch, 80$\, \mu$m width covering an arc for $\approx 90\,^{\circ}$) for the measurement of the radial coordinate and a matrix of 1560 pads, each one covering $\Delta\eta\times\Delta\phi \approx 0.06 \times 0.018$ rad, for the measurement of the azimuthal coordinate and for triggering.

The pad and strips signals (2072 per detector plane) are available in digital form and only the list of the active pads and strips is saved.

Radial and azimuthal coordinate resolution is about $110 \, \mu$m and $1 ^\circ$, respectively. The total material of 10 GEM detectors amounts only to $\approx 0.05 \, X_o$.

\subsection{Track Reconstruction}

The amount of particles produced by the interaction of primary particles with the material in front of and around T2 is particularly challenging both for the detector performance and for the physics analysis.

A special effort was devoted to understand and quantify secondary particles produced by the interaction of particles with the material in front of and around T2 and then detected in the telescope.

The simulation of the forward region, properly tuned with the data, showed that a large number of secondary particles (roughly 90\% of the signal in T2) are produced in the vacuum chamber walls in front of the detector, in the beam pipe (BP) cone at $ | \eta | = 5.53 $ and in the lower edges of the CMS Hadron Forward calorimeter (HF).\cite{T2:Geant4-T2}

These are responsible for most of the high multiplicity events in T2 and produce a strip occupancy larger than 40\% in $\sim10\%$ of the events.

Besides the remnant magnetic field, which is locally weak and almost aligned with the track direction, does not provide any selection possibility for the lowest energy particles.

Specific aspects of the T2 analysis, that proceeds within the software framework described before, are outlined in what follows.

The GEM signal digitisation is parameterised following a detailed simulation that reproduces well the behavior of the detector in terms of the cluster size (i.e. a group of hits in neighboring strips or pads) and the reconstruction efficiency as a function of the ionisation energy released in the gas by the incident particle (with proper consideration of diffusion coefficients of the gas mixture, detector gain and the VFAT amplifier thresholds).

\begin{figure}[!h]
\begin{center}
\includegraphics[width=0.98\linewidth]{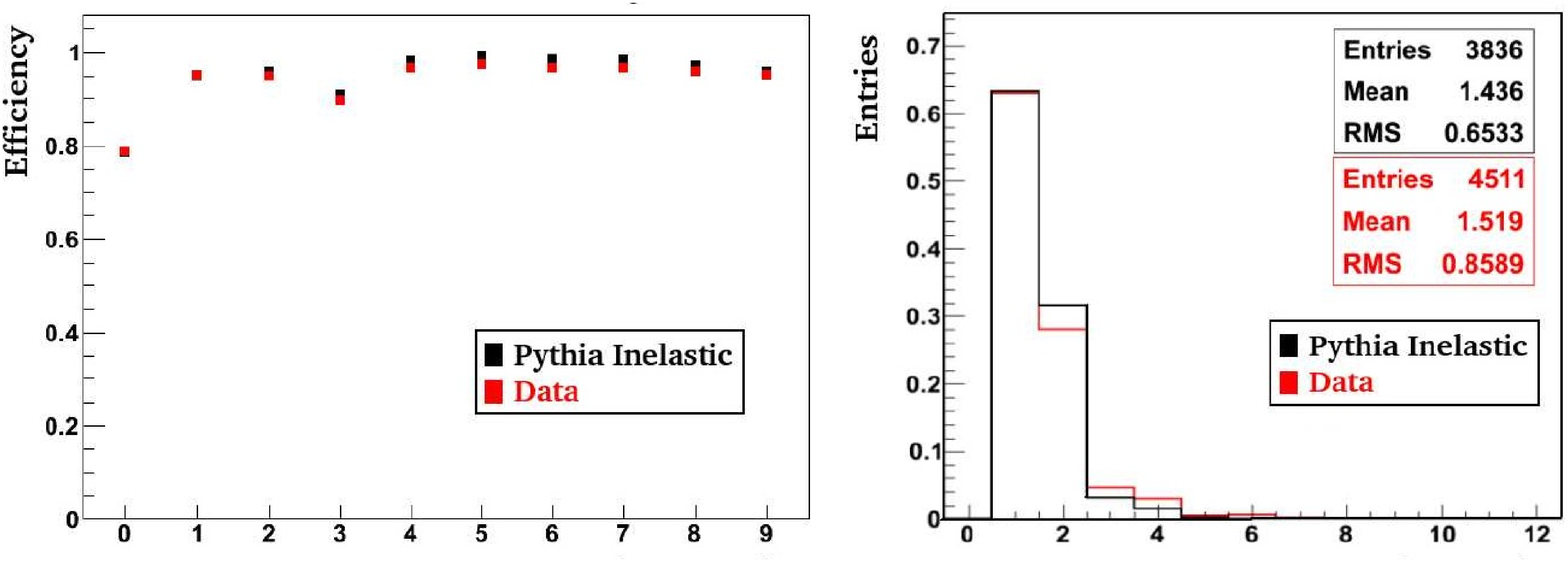}
\caption{Comparison of detector behavior for 7 TeV data and MC (Pythia) inelastic events: plane pad efficiency (left) and cumulative pad cluster size (right) for one of the T2 quarter. Results of similar quality have been obtained for the strips.}
 \label{fig:T2-PlanePadEffiCLS_H1}
\end{center}
\end{figure}

The results on the comparison between data and tuned simulation for the pad efficiency and cluster size after taking into account  the fraction of dead or noisy channels, measured to be ~6\% for the pads and ~9.5\% for the strips, is shown in Fig. \ref{fig:T2-PlanePadEffiCLS_H1}. 
Similar results are obtained for the signals generated by the strips.

The track reconstruction is based on a Kalman Filter-like algorithm\cite{MirkoTesi}, simplified due to the small amount of material traversed by the particle crossing the 10 GEM planes and to the low local magnetic field in the T2 region.

The particle trajectory can, therefore, be successfully reconstructed using a straight line fit.

The minimum requirement for a straight line fit is 4 hits (pad clusters with or without an overlapping strip cluster), of which at least 3 had a pad/strip cluster overlap. A $\chi^{2}$-probability greater than $1\%$ is required for the straight line fit.

The $\eta$-value of a track is defined as the average pseudorapidity of the T2 track hits, calculated from the angle that the hit has with respect to the beam at the IP. The definition has been adopted on the basis of detailed MC simulation studies to find the optimal definition of the true $\eta$ of a particle produced at the IP. 
With this definition of the track $\eta$, a resolution better than 0.04 units is obtained at the center of the pseudorapidity acceptance of T2.

\subsubsection{Primary track selection}\label{section:primtracksel}

The coordinate system defined for the track reconstruction has the origin located at the nominal collision point, the X axis pointing towards the centre of the LHC ring, the Y axis pointing upward (perpendicular to the LHC plane), and the Z axis along the counterclockwise beam direction.

\begin{figure}[htb]
\centering
\includegraphics[width=0.49\linewidth]{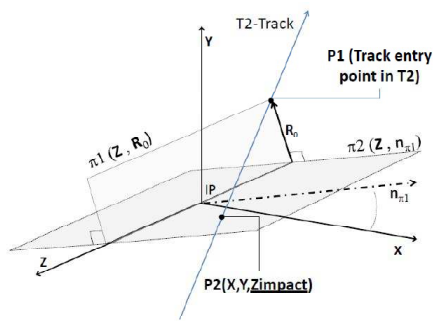}
\caption{Definition of the track Z$_{Impact}$ parameter used for primary track selection.}
\label{MCPrimSecfitsBeforeAftera}
\end{figure}

Two geometrical parameters Z$_{0}$ and Z$_{Impact}$ are used to describe a track that is detected in T2, where Z$_{0}$ is the Z value at the point of minimum approach of the track to the Z axis and Z$_{Impact}$  is the Z coordinate of the intersection between the track and a plane (``$\pi_{2}$'') containing the Z axis and orthogonal to another plane defined by the Z axis and the track entry point in T2 (``$\pi_{1}$'')(see Fig. \ref{MCPrimSecfitsBeforeAftera}).
Due to the short lever arm provided by the T2 detector ($\sim\,$40 cm), if compared to the distance to the IP ($\sim\,$13.5 m), the Z$_{Impact}$ and Z$_{0}$ resolution are both typically larger than 1 m. 

Since about 80\% of the T2 reconstructed tracks are secondaries, it is important to define an appropriate procedure for the discrimination between tracks generated by primary and secondary charged particles.

\begin{figure}[htb]
\centering
\includegraphics[width=0.6\linewidth]{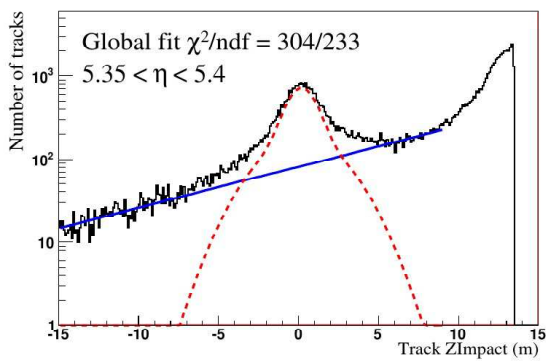}
\caption{The track Z$_{Impact}$ parameter distribution, obtained in one quarter of the T2 detector for tracks with $5.35<|\eta|<5.40$ . The $\chi^2$/ndf =281/226 refers to the global (double Gaussian + exponential) fit performed for z ranging from -15 m to 9 m. }
\label{fig:Zimpact}
\end{figure}

Based on detailed simulation studies, the most effective primary/secondary particle separation is achieved using the Z$_{Impact}$ track parameter\cite{MirkoTesi}. This parameter is stable for misalignment errors and is well described by a double Gaussian function for the primary particles and by an exponential function for the secondaries.
The track Z$_{Impact}$ distribution for  tracks reconstructed from data in one T2 quarter in the $5.35<|\eta|<5.40$ range and the exponential and double Gaussian fit is shown in Fig.~\ref{fig:Zimpact}. The blue-solid curve represents the exponential component due to secondaries, while the red-dashed curve is the double Gaussian component mainly related to primary tracks. 
Similar results are obtained for ranges of $\eta$ that cover the acceptance of the detector.

\begin{figure}[hbt]
\centering
\includegraphics[width=0.6\linewidth]{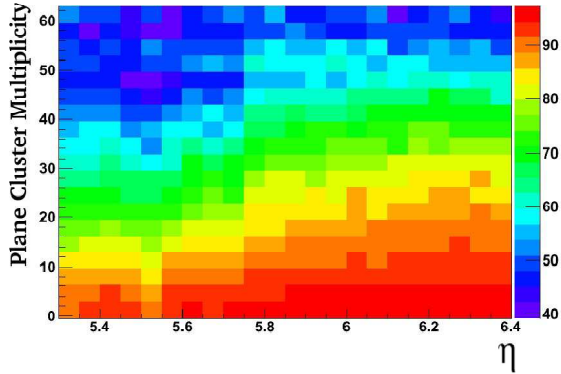}
\caption{Primary track efficiency after applying cuts in $Z_0$ and $Z_{Impact}$ (see description in the text).}
\label{fig:T2-2D-primary-eff}
\end{figure}

The requirement that {$Z_{0}\, \cdot\, {\rm sign}(\eta)\, < 13.5\,$m} reduces the amount of secondary tracks by about 60\%. 
The additional requirement that also the track Z$_{Impact}$ be contained in the z range for which the double gaussian area contains 96\% of the tracks,  gives a  primary track selection purity of better than 80\%.
The primary track efficiency, evaluated with a simulation, is defined as the probability to successfully reconstruct a GEANT4 generated primary track that traverses the detector with the $Z_0$ and $Z_{Impact}$ parameters within the range described above.
The efficiency has been found to depend both on $\eta$ and on the detector occupancy (see Fig.~\ref{fig:T2-2D-primary-eff}). 
Applying the combined $Z_0$ and $Z_{Impact}$ requirement an average primary track efficiency between $\approx 75\%$ to $\approx 85\%$ is obtained.

A single track efficiency greater than $98\%$ is obtained if one does not impose any cuts on the track parameters.

\subsubsection{Alignment}

The relative position of the detector planes within a T2 quarter (internal alignment) and the overall alignment of the detector planes with respect to their nominal position (global alignment) have been investigated in detail to define possible misalignment biases of the track measurements\cite{MirkoTesi}.

The shifts of the planes in the directions transverse to the beam ( X and Y) are the most important internal alignment parameters. 
Two different methods were used to correct for such displacements and both gave consistent results, with an uncertainty on the transverse position of the plane of about 10$\,\mu$m.
The relative alignment between the two neighboring quarters was obtained using tracks reconstructed in the overlap region.

The global alignment of the detector is of main importance for the analysis and is obtained by exploiting the symmetric distribution of the track parameters and the position of the ``shadow'', a circular shaped zone of the T2 planes characterised by a very low hit rate, due to interactions of primary particles with the thick beam pipe flange in front of T2 at $|\eta|\,=\,$5.53.

The X-Y shifts with respect to the nominal position and the tilts in the XZ and YZ planes are determined with a precision respectively of $\sim\,$1 mm and of $\sim\, 0.3\div0.4$ mrad.

\begin{figure}[htb]
\centering
\includegraphics[width=0.98\linewidth]{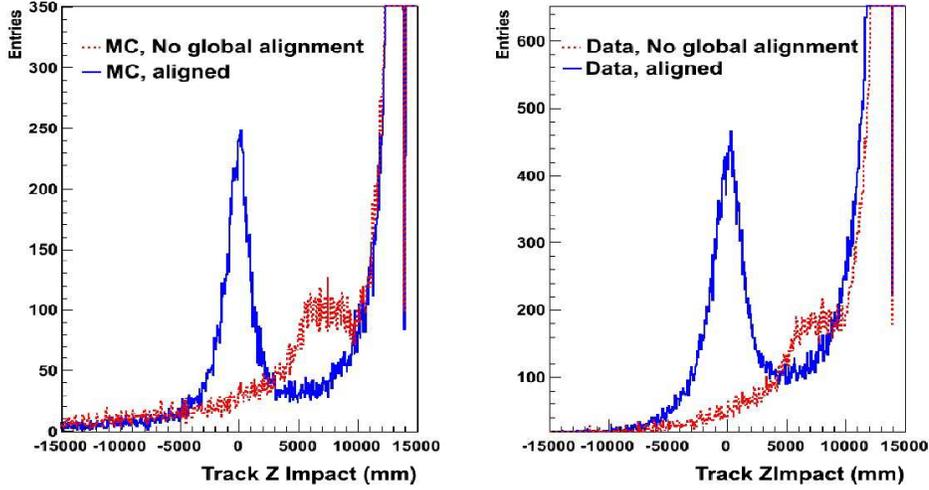}
\caption{Track Z$_{Impact}$ parameters (see secition\ref{section:primtracksel} for its definition) reconstructed in one of the T2 quarter (left plot from simulation and right from data). The blue (red) curve is obtained with (without) the inclusion of the global alignment corrections.}
\label{fig:Zimpact-distr}

\end{figure}

The measured local and global alignment parameters of the telescope are introduced into the GEANT4 simulation and the algorithm for the correction of the hit positions applied to the reconstruction of both simulation and data.
The Z$_{Impact}$  parameter is very important for the selection of the primary tracks for an aligned quarter, and should be symmetric around Z$_{Impact}$=0.
This track parameters is very sensitive to the misalignment as can be inferred from  Fig.~\ref{fig:Zimpact-distr} that shows the distribution of the track Z$_{Impact}$ parameter for both MC and data events with and without global alignment corrections.

As shown in the figure the primary track signature would be completely lost without a proper global alignment correction.

\subsubsection{Acceptance}

The T2 telescopes detects a very large fraction ($\approx 94\%$) of the inelastic cross section and in particular of the diffractive processes with small masses of the diffractive system ($M_{diff}$), where particles are produced at very small angle with respect to the beam. 
The $(dN/dM_{diff})$ distribution is expected to peak at masses of 1-2 GeV, and the acceptance is smoothly varying from 0\% to 100\% from $\approx 2\,$GeV to $10\,$GeV. 
The T2 acceptance edge of $|\eta|$ = 6.5 corresponds to a diffractive mass of about 3.4 GeV (at 50\% efficiency), hence the majority of the events with mass below 3 GeV will not be detected.

\begin{figure}[b]
\centering
\includegraphics[width=0.55\linewidth]{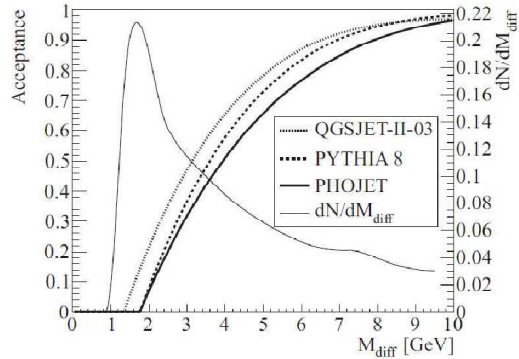}
\caption{The acceptance of the combined T1 and T2 detectors as a function of the diffractive mass and the diffractive mass distribution $(dN/dM_{diff})$ for single diffractive events.}
\label{fig:T2-T1-acceptance}
\end{figure}

The diffractive mass $(dN/dM_{diff})$ distribution has been simulated with QGSJET-II-03 and is shown in Fig.~\ref{fig:T2-T1-acceptance} together with the combined T1 and T2 mass acceptance simulated with PYTHIA8, PHOJET and QGSJET-II-03 as a function of $M_{diff}$.

The transverse momentum ($p_{T}$) acceptance for single charged particles detected by T2 is limited by the magnetic field and multiple scattering effects. Simulation studies have shown that the charged particle tracks are reconstructed with a good efficiency for $p_{T}\geq\,$40 MeV/c, defining effectively the minimum $p_{T}$ acceptance.

\begin{figure}[thb]
\centering
\includegraphics[width=0.55\linewidth]{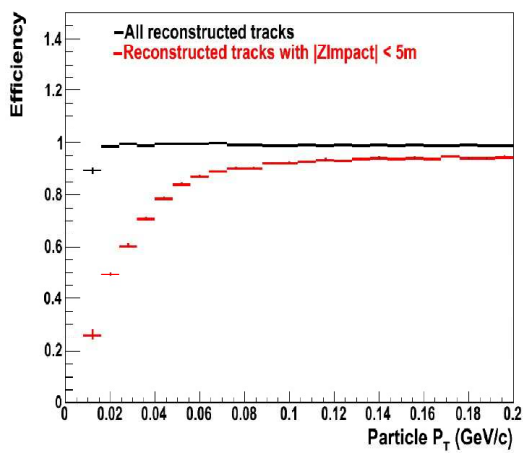}
\caption{Efficiency of single pion reconstruction as a function of the particle $p_T$. The red and black curves show respectively the $p_T$ acceptance obtained with and without applying the primary selection cut on the Z$_{Impact}$ parameter.}
\label{fig:T2-acceptance}
\end{figure}

The efficiency of reconstruction for tracks of a given p$_T$ traversing  the T2 telescope is shown in Fig.~\ref{fig:T2-acceptance}.
The fraction of charged particles with $p_{T}<\,$40 MeV/c produced in the T2 acceptance is predicted to be very small ($\sim\,$1\%).

\subsection{The minimum bias trigger generation for T2}

Also in the case of T2 the fast trigger signals in the VFAT R/O generated by the passage of a particle in the T2 telescope are analyzed locally on the detectors via a majority coincidence unit, the Coincidence Chip (CC).

Pads on each plane are grouped in the CC to form {\it Superpads} hit\footnote{A 'Superpad' is a sector of 5x3 pads grouped together, see section~\ref{T2-section}}.

A "trigger road", defined as the Superpads in corresponding $r-\phi$ sectors of the 10 planes in the same T2 quarter, signal is generated if a number of superpads in the road are hit: the number of planes required for a "Trigger Road" can be varied and is usually defined as $\ge 3$ SuperPads in a road.

To generate a T2 trigger at least one trigger road must be set in one of the 4 quarters, condition that is satisfied if at least one charged particle traverses the T2 detector.
The efficiency of the trigger generation is checked regularly and it is calculated with three methods comparing the online track reconstruction algorithm (trigger road), the offline track reconstruction algorithm, and the primary tracks vertex reconstruction.

An average T2 trigger efficiency of ($98\pm1$)\% was measured, with little difference among the three methods.\cite{Aspell:2012ux,PhysRevLett.111.012001}

\section{The T1 telescope}

The T1 telescope of the TOTEM experiment complements T2 in the measurement of the inelastic rate of proton-proton interactions at the LHC over an angular range from 3.1 $<\vert \eta \vert< $ 4.7  and is installed in the forward cone of the CMS end caps where a non uniform and poorly known magnetic field is still present.

Each arm of T1 is composed of five planes of Cathode Strip Chambers (CSC), with six detectors per plane each covering roughly a region of $60^ \circ $ in $\phi$. One arm is split in two halves mounted on two different supports. 
The five planes of detectors in a telescope are rotated with respect to each other of $5.6^{\circ}$ per plane, a feature that is useful for pattern recognition and helps to min minimize detector overlaps  localized concentration of material in front of the CMS HF (Hadronic Forward) Calorimeter.

Two stiff honeycomb panels of trapezoidal shape determine the flat surfaces of the CSC's cathode planes.
The cathode electrodes are parallel strips obtained as gold-plated electrodes oriented at$\pm 60^o$ with respect to the direction of the anode wires and have 5.0 mm pitch (4.5 mm width and 0.5 mm separation). 
The anode plane of the detector consists of gold-plated tungsten wires of $30\mu$m diameter and a pitch of 3mm. 
This electrodes configuration allows three coordinate measurements in one plane for each particle track, which significantly helps in resolving multiple events. 
The front-end electronics card is directly soldered onto pads connected to each single anode wire via decoupling capacitors and protection diodes.

A detailed description of the T1 telescope can be found in  [\citen{Anelli:2008zza}].

\subsection{Track reconstruction}

The T1 data analysis uses the same analysis framework described before and is described briefly in what follows.

\begin{figure}[ht]
\centering
\includegraphics[width=0.98\linewidth]{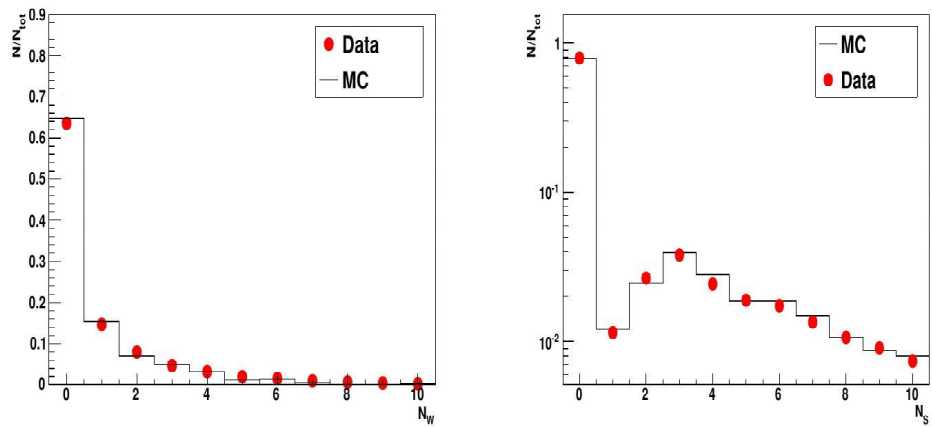}
\caption{Comparison between MC and data for the T1 CSC detector simulation, left plot shows the number of anode wires with signal per each event, right plot the same for the cathode strips.}
\label{fig:T1-MC-data-detector}
\end{figure}

 A specific digitization module has been developed for the T1 simulation. It simulates 
the total charge deposited on one wire with the application of the Gatti function\cite{gatti:function} and provides an evaluation of the signal on the cathode strips and the percentage of the anodic charge induced on the cathode planes.
A bit for the relevant anode wire or cathode strip (or strips) is set after proper consideration of threshold and noise of the signal amplifier.
The cathodes cluster parameters (geometrical center and half width) are calculated if the induced charge brings above threshold at least two adjacent strips.
In Fig.~\ref{fig:T1-MC-data-detector} Data-MC comparisons show that the detector response is properly implemented in the simulation.

The T1 track reconstruction is done in two steps.
A pattern recognition algorithm associates hits, defined as a weighted triple coincidence of anode and cathodes signals of one chamber, with regions of interest (roads)  $\Delta \eta \times \Delta \phi = 0.1 \times 0.2 $ radians that have a good probability of containing a track coming from the interaction point. 
A track is fitted from the hits if the road includes at least 4 and up to 30 hits distributed in at least 3 or more of the 5 planes.
The number of hits and road size had been modeled with a MC simulation to optimize both tracking efficiency and computing time. 

A track fitting algorithm, a straight line fitting independently the xz and yz projection, is then applied to each road.
Even in the presence of a non uniform and poorly known magnetic field in the T1 region where the lines bend to enter the return iron yoke of the CMS solenoid, a straight line approximation has been chosen: MC studies in fact have shown that the straight line approximation works properly in the present analysis. 

\begin{figure}[ht]
\centering
\includegraphics[width=0.55\linewidth]{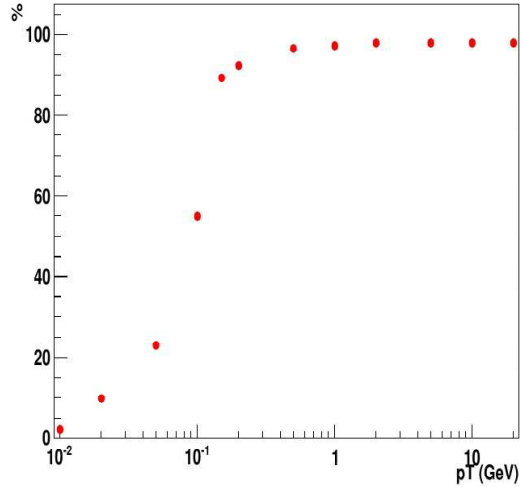}
\caption{The $p_T$ acceptance of the T1 telescope.}
\label{fig:T1-PT-acceptance}
\end{figure}

The plot in Fig.~\ref{fig:T1-PT-acceptance} shows the acceptance of simulated primary particles as a function of $p_T$ and can be seen that the algorithm successfully reconstruct trajectories of particles with transverse momentum $p_T \ge 100$ MeV.  

\begin{figure}[hbt]
\centering
\includegraphics[width=0.55\linewidth]{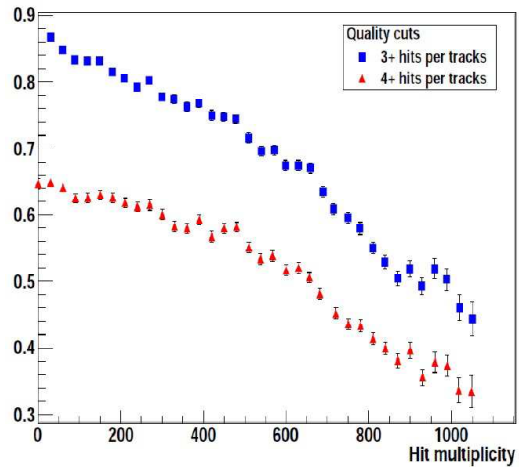}
\caption{The simulated efficiency of primary tracks reconstruction: the red triangle points have an added quality requirement of 4 points per track, while the blue square ones define a track only with three points (see text).}
\label{fig:T1-reco-eff-hits}
\end{figure}

The primary tracks reconstruction efficiency depends also on the multiplicities of the reconstructed hits in each detector where primary track hits get mixed with secondaries  and ghost hits .
The dependence of the probability for a primary track to be reconstructed once it arrives in T1  for two different quality requirements for the track reconstruction is shown in Fig.~\ref{fig:T1-reco-eff-hits}.

The red line in the plot shows a decrease in track efficiency when one adds a more stringent track requirement of at least four points per track.

\begin{figure}[ht] 
\centering
\includegraphics[width=0.98\linewidth]{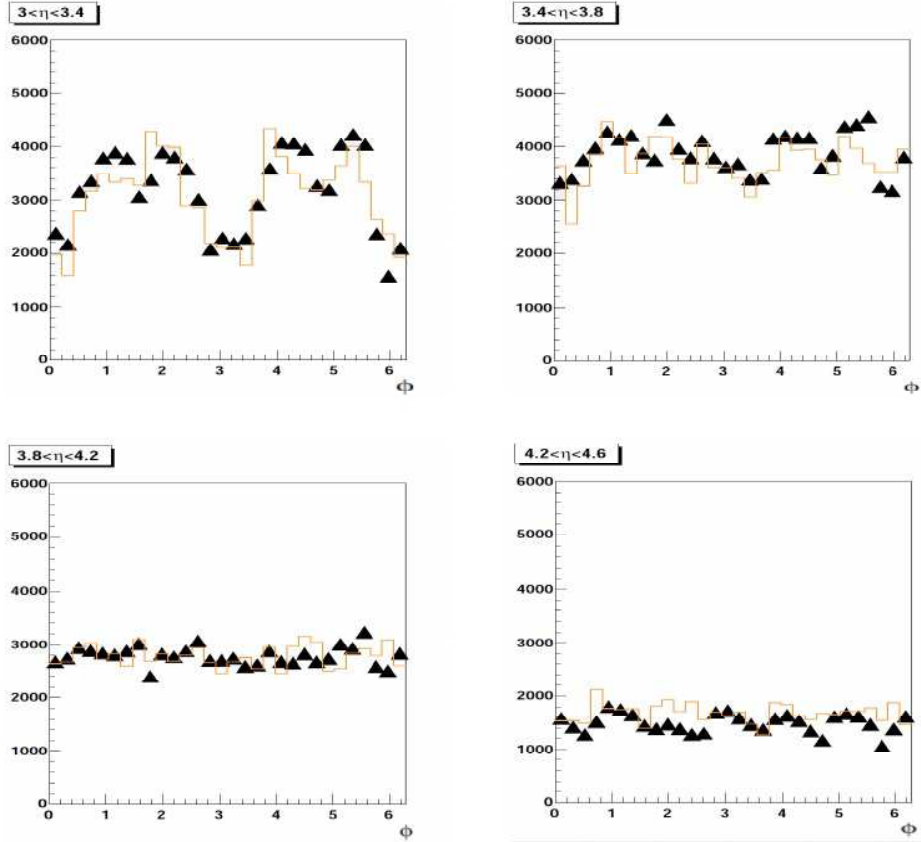}
\caption{T1 $\phi$ tracks distribution for 4 different $\eta$ intervals  (the orange continuous line represent the result of the MC simulation, data points are indicated with triangles).}
\label{fig:T1-phi-dist}
\end{figure}

The  $\eta$  and $\phi$ distribution of reconstructed tracks, mainly affected by chamber inefficiencies and geometrical coverage, is also well reproduced by the simulations. 
The tracks from charged primary  in T1 are reconstructed with an efficiency which depends on the hit multiplicity, with a maximum of $\approx 87$\% for very low multiplicity events.
 Fig.~\ref{fig:T1-phi-dist} shows the $\phi$ distribution for reconstructed tracks in T1 for different $\eta$ intervals.
The non uniformities in the distribution for the low $\eta $ range in the regions at $\phi =0$ and $\pi$ are due to the smaller radial dimensions of the detectors.  

The resolution with which the pseudorapidity is measured depends on the particle $p_T$ mainly because of the effect of the magnetic field, which is far from negligible both in front and inside T1, and ranges from $\approx$1\% to $\approx$10\%.
The single track event reconstruction efficiency is estimated to be $\approx$98\% using simulation with data tuned CSC efficiencies.	

\subsubsection{Alignment}

The T1 detectors alignment is a three step process using a sample of reconstructed hits and tracks from the data.

\begin{figure}[!hbt]
\centering
\includegraphics[width=0.49\linewidth]{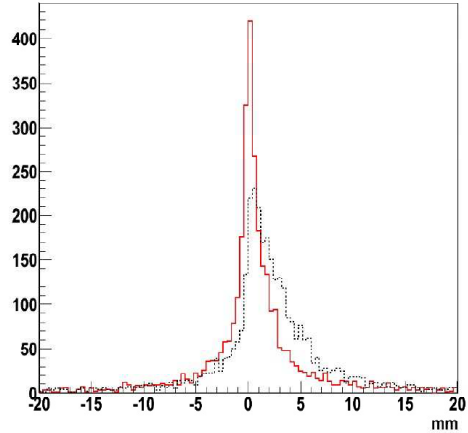}
\caption{Residuals for the T1 detectors before and after the alignment procedure: the black dotted line is the distribution of the values  before, the red continuous line after the alignment operation.}
\label{fig:T1-detector-alignment}
\end{figure}

The alignment of the transverse coordinates of the chambers in the 5 planes of each T1 sextant with respect to each other is provided by a $\chi^2$ minimization.
In order to minimize the distance between the reconstructed hit and the intersection of the track with the chamber plane one iterates a 15 parameters fit  ($\Delta x_i, \Delta y_i,\Delta \phi_i$) applying a roto-translation per chamber of the sextant.
The 6 sextants of each T1 arm are then aligned with respect to each other using tracks reconstructed in the overlap regions between adjacent sextants. 
The residuals of one detector before and after the alignment procedure are reproduced in Fig.~\ref{fig:T1-detector-alignment}.

\begin{figure}[!hbt]
\centering
\includegraphics[width=0.98\linewidth]{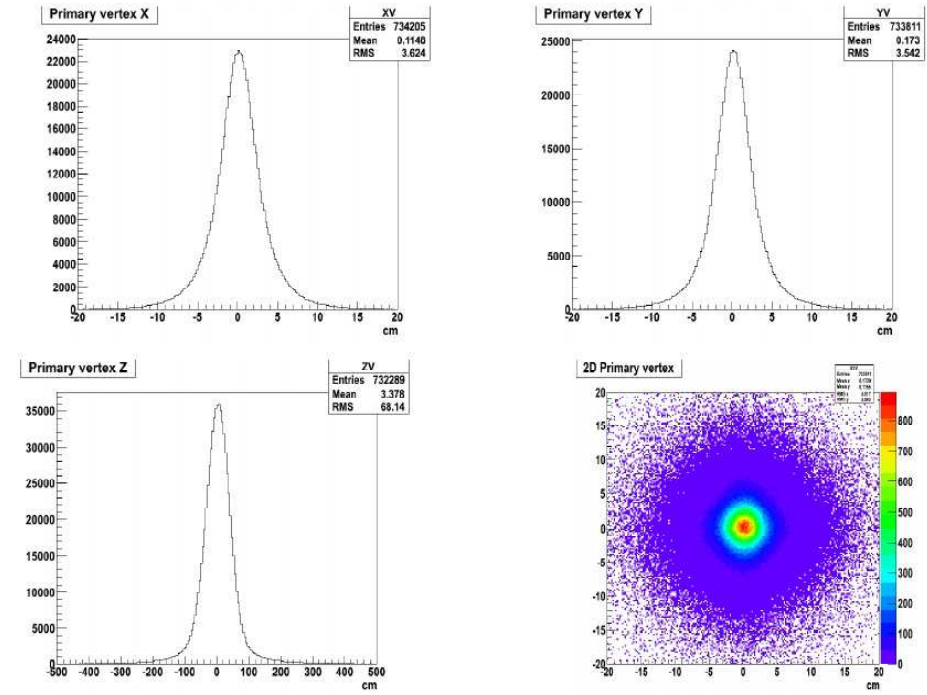}
\caption{Vertex position distribution as reconstructed with the T1 data.}\label{fig:T1-vertex-alignment}
\end{figure}

After the alignment the position of each chamber is known with an uncertainty of $\approx$ 3.9 mm  in z and $< 100 \, \mu$m the xy plane.
The final step aligns the two arms with respect to each other and to the beam using the capability of T1 to reconstruct the three coordinates of the primary vertex (Fig.~\ref{fig:T1-vertex-alignment}).
Alignment is performed with low multiplicity events of less than 10 tracks to avoid possible side effects due to crowded events.

In order to estimate the systematic error  induced on the track $\eta$ by the alignment one assumes an 100\% error on the corrections for high rapidity tracks, the most affected from a tracking error which is uniform over the detector area. 
The induced error on $\eta$ is less than 1\%, negligible or at most comparable to the systematic error due to multiple scattering and magnetic field effects.

\subsubsection{Vertex reconstruction}

A vertex finding algorithm has been implemented to measure the primary vertex coordinates. The primary vertex position as measured in real data is shown in Fig.~\ref{fig:T1-vertex-alignment}.

The RMS in the transverse coordinates is comparable with the beam pipe size (~3cm) and in the longitudinal direction with the interaction region size (~50cm).

\subsection{Inelastic rate: event counting efficiency}

An event is missed by T1 if no track is reconstructed for the event.
In the MC sample of 10,000 events the number of missed events is  649 considering a fully efficient detector and increases to  657 after correcting for the estimated local inefficiencies.
The probability that T1 will be unable to identify an event  is 0.08\% .

\subsection{Minimum bias trigger generation for T1}

The T1 detectors'  plane are distributed over a  distance of almost 3m and rotated with respect to each other and this has suggested that the raw trigger hits generated by the VFAT fast outputs be sent directly to the counting room to be analysed centrally.

\begin{figure}[hbt!]
\centering
\includegraphics[width=0.7\linewidth]{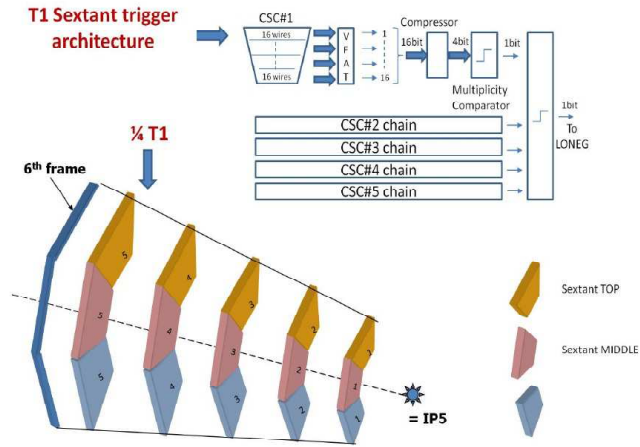}
\caption{T1-sextant trigger logic implementation and representation of a T1 sextant.}
\label{fig:T1sextant}
\end{figure}

The  trigger information is provided by the CSC anode wires, and the trigger bits of the anodes of a half plane (3 detectors) are grouped locally on the Read-Out card (ROC \cite{Minutoli:2010zz}),  serialized and optically transmitted to the T1 TOTFED in the service Cavern where, after a plane by plane  “Hit count” (bit set if the hits multiplicity for the plane is larger than a preset value  $N_{mult}$),  the multiplicities from each plane of the same $\phi$ sextant\footnote{The CSC detectors  on the  5 planes of T1 that cover approximately the same $\phi$ range form a sextant.} are compared  and a global sextant trigger bit is generated by a majority logic (Fig. \ref{fig:T1sextant}). 

A T1-trigger bit signal and the 12 sextant trigger bits are then coded in a T1 trigger word to be analysed with the other TOTEM detector signals.

\section{Triggering TOTEM}\label{trigger-section}

Using the trigger information collected from the different sub-detectors (physically separate and different in design) along with the need for distinct algorithms for the selection of elastic and inelastic events require a trigger system offering a wide range of possibilities.

\begin{figure}[!hbt]
\centering
\includegraphics[width=0.8\linewidth]{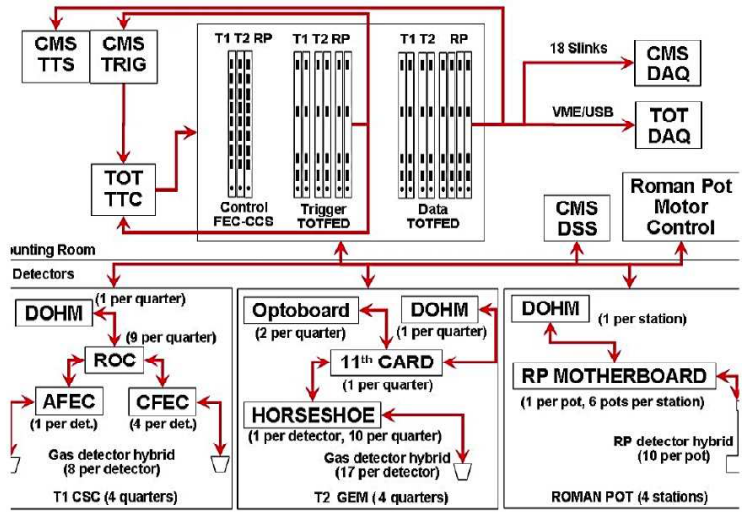}
\caption{The TOTEM trigger logic diagram.}
\label{fig:trigger-logic}
\end{figure}

The TOTEM  trigger system is divided in three substructures: the first at detector level generates fast trigger signals, the second, after the transmission to the counting room, makes initial association detector by detector, and the third  puts all the information  together to generate the final trigger.
The complete trigger architecture is shown in figure \ref{fig:trigger-logic}.

The initial fast trigger information is generated for all the 3 detectors as the OR of hits in 16 adjacent VFAT channels in the VFAT2 ASIC, and transmitted through the fast bits bus. 
These Low resolution trigger hits are available on the  8 bit wide bus one clock cycle after the event.
Various configurations of the fast bits can be defined/selected through an I2C register.
This information is sent to the central trigger in the counting room through optical links by a GOH hybrid that transmits serially 16 bits per fiber to the OPTORXs of the trigger boards which then deserialise them.

In view of the possibility of using both the CMS and the TOTEM detector to study forward physics   the trigger is designed to be compatible with CMS
and special attention has to be paid to the arrival time in the counting room.
In consideration of this the transmission of the signals from the Roman Pots located at 220m from the interaction point needs to be treated in a special way.
Considering that the signals in the RP detectors are generated with a delay respect to the interaction time due to the 220m of flight of the protons from IP5 to the RP , to transmit them by optical fiber as for the rest of the experiment and then would have to undergo the electronics transit setup time (serializer and deserializer operations) and considering the latency due to the required optical fiber length, these would reach the counting room after a latency incompatible with the CMS latency.
The trigger bits from the RP are instead sent via special coaxial cables with a propagation delay lower than the optical fiber (4.2 ns per meter in the coaxial cable as compared to 5.0 ns per meter in the optical fiber). Given the length of the transmission line of $\approx 270$m a parallel LVDS bus  with LVDS repeaters on the path has been developed.
This solution while preserving good timing information allows to reduce by 20 clock cycle (~500 ns) the overall response of the trigger signals which now become compatible with the CMS maximum latency.

As mentioned the RPs and T2 Trigger Hits are analyzed locally on the detectors via a majority coincidence unit, for T1 the raw trigger hits are sent directly to the counting room to be analysed centrally.

The second trigger step is performed in the OPTORX/Main FPGAs  hosted in the TotFed boards. 
In the RP, T1 and T2 TotFed  the trigger roads are counted and ORed to form a coded trigger word containing all the information for the event of that specific detector. 

The third step is performed in a special TOTEM board called LOneG (Level One Generator) \cite{bagliesi:loneg} where a single FPGA, Altera StratixII, puts together the information from all the detectors. 
The first operation is aligning in time all the information coming from the detectors that are distributed over a distance of 440m along the beam line. 
Then these trigger data are input to a Look Up Table which generates trigger bits for up to 16 different trigger combinations. 
And as last operation each trigger combination bit is put in coincidence with a bunch scheme strobe (“fork” strobe) in order to select a particular trigger combination only on well defined bunch crossings.

A prescaling factor can be applied to any trigger bit in either path.
Beam crossing trigger may also be generated in the LOneG to provide a way to estimate the trigger inefficiencies and the pile-up probability. 
An appropriate prescaling factor is applied to the rate of the bunch crossing events to keep the overall trigger rate around 650Hz, well below the DAQ rate limit of 1 kHz.

The Totem L1 is the OR of all the  triggers  enabled amongst the up to 16 different algorithms that may be implemented by the LONEG. 

In order to be able to send to CMS ( and receive from) triggers signals with the appropriate timing  the last stage is split into 2 parallel paths for which one may apply different “fork” strobes and may define a different programmable final latency. 
The first one generates the trigger pattern sent to the CMS L1 trigger, the second one generates the Totem trigger. 
To complete the description of the logic implemented in the LOneG the second path receives in input two signals from CMS, the CMS L1 signal and the TOTEM special trigger bit as generated by CMS L1 algorithms (L1SA). 

The trigger information created for every TOTEM L1is written into a Daq frame after passing the trigger throttling rules by the LTC board . 
The Trigger frame thus contains, event number, Trigger pattern, Trigger number (Number of trigger generated before the trigger throttling suppression in the LTC board), Bunch number, Orbit number and, if available, other information for event reconstruction.
The overall latency of TOTEM L1 trigger depends on the loop TOTEM-CMS-TOTEM and is 150 clock cycles. The CMS path latency is set around 96 clock cycles.

\subsection{Sharing trigger and data information with CMS}\label{triggering:cms}

The Trigger (and data) synchronization between CMS and Totem is achieved by counting  for each event the Bunch Number in each orbit and the Orbit Number from orbit counters. 
\begin{figure}[h]
\begin{center}
\includegraphics[width=0.98\linewidth]{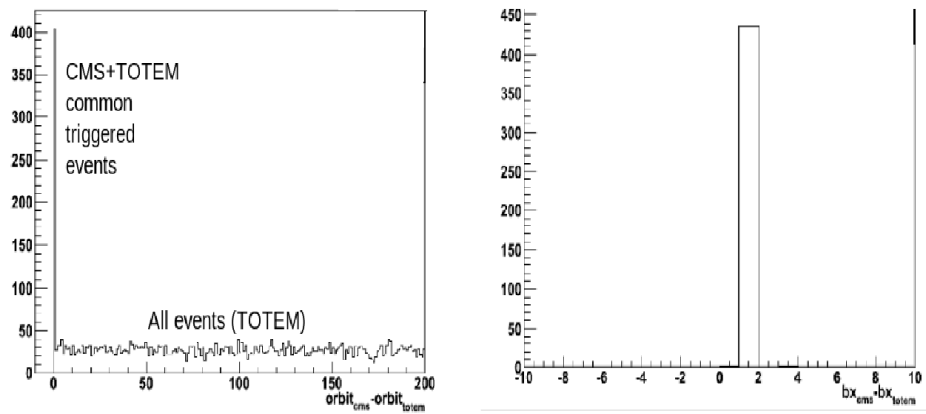}
\caption{Distribution of offset of the orbit number (left)  and of the bunch number (right) for a reconstructed run.  }
\label{fig:offset}
\end{center}
\end{figure}
The Orbit number counters count the orbits synchronously in both experiments TOTEM and CMS. These counters are synchronised (reset) for both experiments by  CMS at the beginning of each of their runs.

Data synchronization between CMS and Totem is then achieved offline on the reconstructed data from the common runs by aligning, for each event , the Orbit Number from "orbit counters" and the Bunch Number in each orbit.
The two sets of reconstructed data are then merged according to the synchronization selection.
Fig~\ref{fig:offset} shows results of the merging procedure for one run: the bunch number offset between the two experiment is equal to 1 due to a different counting convention in the two experiments and the orbit offset for the events which have been triggered by the same condition is equal to 0.

\subsection{Determination of the machine luminosity}

Once the cross-section of a process is known, it can be, in general, used for luminosity determination. The TOTEM experiment can then measure directly the machine luminosity.

TOTEM can detect elastically scattered protons at very low values of $|t|$: dedicated runs with high $\beta^*$ optics (see Sec.~2.1)have been essential for absolute luminosity determination based on the optical theorem:

\begin{equation}
\mathcal{L} = \frac{1+\rho^2}{16 \pi (\hbar c)^2}\;\;\frac{(\rate{N}{el}+\rate{N}{inel})^2}{\left.\difrate{N}{el}\right|_{t=0}} %
\label{eq:lumi}
\end{equation}
where $\rate{N}{el}$ is the elastic hadronic rate and $\rate{N}{inel}$ is the inelastic one, both integrated over a run period; the quantity $\rho$ stands for ratio of the real to imaginary elastic hadronic amplitude at $t=0$. 

Extrapolation to $t = 0$ of the measured differential cross-section for elastic proton proton scattering as a function of the four-momentum transfer squared, $t$, determines the cross-section $d\sigma_{el}/dt\;|_{t=0}$.
Moreover the inelastic cross-section is also directly measured for the same data set by the T1 and T2 telescopes.
A small Monte-Carlo correction ($< 4\%$) is applied to the measured inelastic cross-section to account for the invisible events in the very forward direction $|\eta|>6.5$, mainly due to low-mass single diffraction. 

The integration of the rates over the data-taking period during which the elastic and inelastic interactions have independently but simultaneously been measured provides also the integrated luminosity $\mathcal{L}^{int}_{TOTEM}$.
The  two uncorrelated methods (from the inelastic and inelastic total rate and from the use of the Optical Theorem) give values of the inelastic cross sections that are in excellent agreement and this confirms the understanding of the systematic uncertainties and corrections applied in both methods.

CMS and TOTEM experiments share the same interaction point and the CMS luminosity measurement has been used  in the early stage of the TOTEM data analysis for the TOTEM data normalization. 
The uncertainty of the CMS luminosity measurement, based on Van der Meer scans\cite{CMS:2010gua,CMS-DP-2011-002,CMS:2011vpa,CMS:2012rua,CMS:2012jza}, has been estimated to approximately 4\% for TOTEM runs. 

Both independent methods of Luminosity determination (the CMS and TOTEM ones) were compared at LHC energy of $\sqrt{s}=7$~TeV during an October 2011 run with $\beta^*=90\:$m. 
The integrated luminosity determined by CMS for the October 2011 data set was $83.7 \pm 3.2\;\mu \text{b}^{-1}$ which may be compared to the value obtained by TOTEM of $82.8 \pm 3.3 \;\mu \text{b}^{-1}$ for the same data set; both methods give a similar uncertainty at the level of 4\%, see \cite{totem6} for more details.

\section{Radiation received by the detectors}

In TOTEM, the Total Ionising Dose (TID) and the 1-MeV neutron equivalent particle fluence ($\Phi_{eq}$) are monitored “on-line” during operation using an active Radiation Monitoring (RadMon) system. 
This system consists of sets of RadFET and p-i-n diode sensors distributed across the volume of the three TOTEM sub-detectors. 
\begin{figure}[!ht]
\centering
\includegraphics[width=0.98\linewidth]{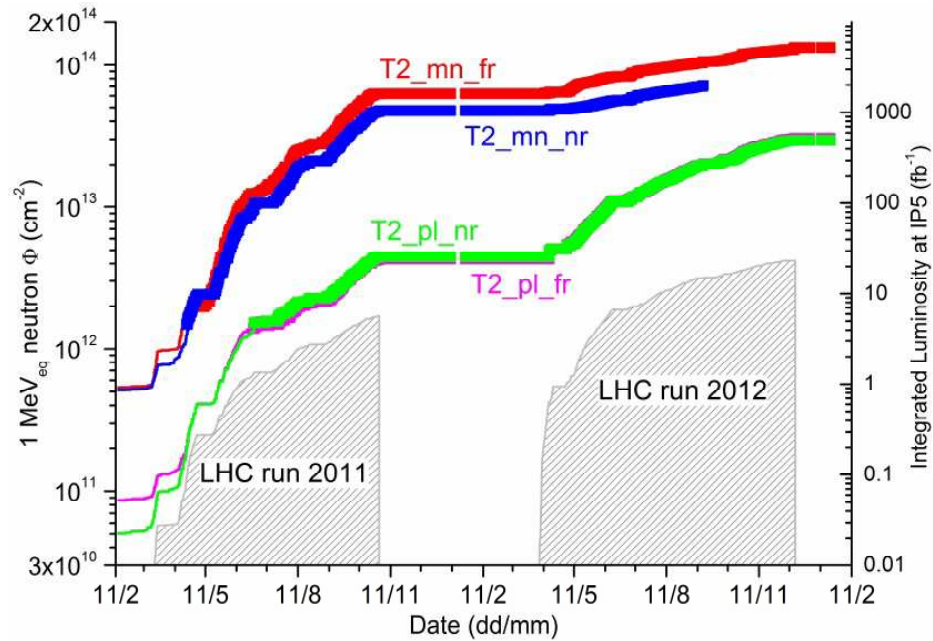}
\caption{Cumulated 1 MeV-neutron equivalent fluence measured with the p-i-n diode sensors: CMRP in the low fluence range (continuous lines), and BPW in the high-fluence range (square markers) for the TOTEM T2 Telescope during the first 3 years of LHC runs.  The evolution of the integrated luminosity  delivered in IP5 during 2011 and 2012 is added for comparison (the gray hashed region).}
\label{fig:dose}
\end{figure}
Two different RadFET (LAAS/REM) and p-i-n diode (CMRP/BPW34) types are needed to provide the required sensitivity and, at the same time, to cover the wide measurement range \cite{Ravotti:sensor}. 
Details about the implementation of the TOTEM system are available in Ref. \citen{Ravotti:studies}. 

Fig.~\ref{fig:dose}  shows the $\Phi_{eq}$ cumulated during the first 3 years of running of LHC in the T2 telescope, the most exposed TOTEM sub-detector. The evolution of the $\Phi_{eq}$ is measured combining data from both CMRP ($10^{10}-10^{12}\,cm^{-2}$) and BPW ($> 10^{12}\,cm^{-2}$) devices. 
The initial measurements were performed with the high sensitivity CMRP p-i-n diodes, while the monitoring of the high-fluence range is performed now with BPW diodes. 
The intensity of the radiation field in T2 scales with the integrated luminosity delivered in IP5 during the 2011 LHC runs, as can be seen in Fig.~\ref{fig:dose}.

The T2 telescope detectors installed on the “minus” arm received an integrated fluence of more than $6\times 10^{13}\,$cm$^{-2}$ by the end of 2011. 
The one located on the “plus” arm integrated an order of magnitude less fluence ($\approx 5 \times 10^{12} \, cm^{-2}$). 
The difference in fluence between the two arms is due to high-energy neutrons generated by the CASTOR calorimeter installed immediately next to T2 but only on the “minus” side of the CMS experiment during the  2011 pp run\cite{Ravotti:studies} due to the absence of the CASTOR calorimeter in CMS; 
 both T2 arms integrated a $\Phi_{eq}$ of a few $10^{13} {\rm cm}^{-2}$ during the 2012 pp run.

\begin{figure}[!hb]
\centering
\includegraphics[width=0.98\linewidth]{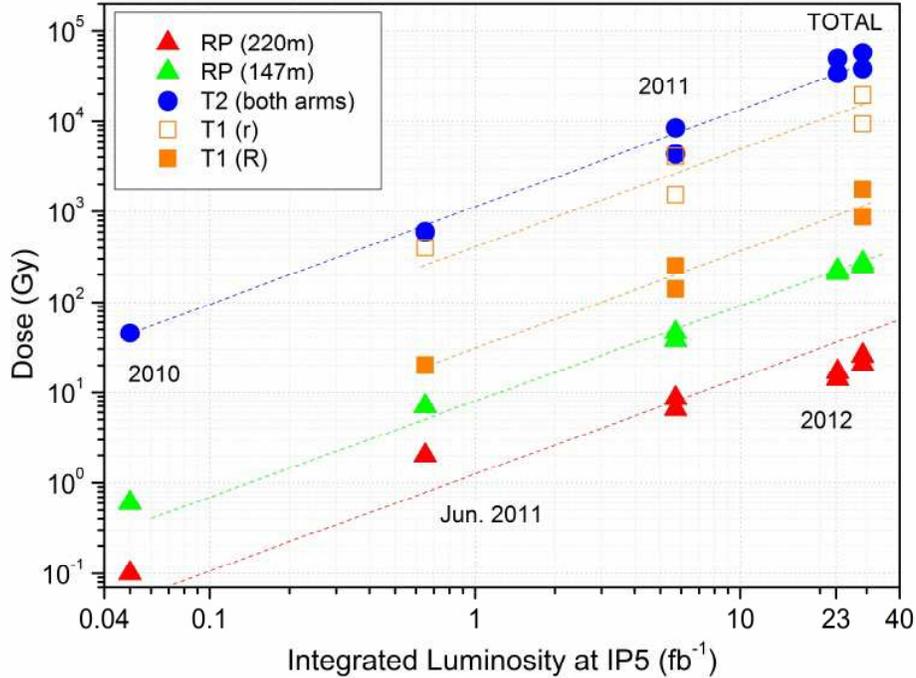}
\caption{Evolution of TID integrated by each of the TOTEM detectors in the first 3 years of LHC runs.}
\label{fig:TID-summary}
\end{figure}
The measured $\Phi_{eq}$  for the large CSC chambers of the T1 Telescope spans about one order of magnitude in the radial direction (transversal plane) with respect to the beam axis. 
Values from $ 2 \times 10^{12} \, {\rm cm}^{-2}$  to   $ 1 \times 10^{13} \, {\rm cm}^{-2}$  have been recorded by the sensors installed on the CSC at large and small radii respectively.

Finally, on the Roman Pot detector stations, a $\Phi_{eq}$ of  $2\times 10^{12}\, {\rm cm}^{-2}$ was measured for the detectors sitting at 147m from IP5, while the cumulated particle fluence on the 220m stations was about 1 order of magnitude lower by the end of the 2011 LHC run.

The evolution of the TID integrated by each of the TOTEM detectors during the first 3 years of running of the LHC is shown in Fig.~\ref{fig:TID-summary}.

\section*{Acknowledgments}

We are grateful to the beam optics development team for the design and the successful commissioning of the high $\beta^∗$ optics and to the LHC machine coordinators for scheduling the dedicated fills. 
We thank P.~Anielski, M. Idzik, I. Jurkowski, P. Kwiecien, R. Lazars, B. Niemczura for their help in software development.
This work was supported by the institutions listed on the front page and partially also by NSF (US), the Magnus Ehrnrooth foundation (Finland), the Waldemar von Frenckell foundation (Finland), the Academy of Finland, the Finnish Academy of Science and Letters (The Vilho, Yrj\"{o} and Kalle V\"{a}is\"{a}l\"{a} Fund), the OTKA grant NK 101438 (Hungary) and the Ch. Simonyi Fund (Hungary).

\bibliographystyle{ws-ijmpa}

\bibliography{performance-bibliography-new}

\end{document}